\documentclass[a4paper, fleqn]{cas-dc}
\usepackage[numbers]{natbib}
\usepackage{amsfonts}
\usepackage{algorithmic}
\usepackage{amsmath} 
\usepackage{amssymb} 
\usepackage{caption}
\usepackage{CJKutf8}
\usepackage{etoolbox}
\usepackage{forest}
\usepackage{lingmacros}
\usepackage{textcomp}
\usepackage{tree-dvips}
\usepackage{tikz}
\usepackage{tikz-cd}
\usepackage[arrowdel]{physics}
\usepackage{graphicx}
\usepackage{wrapfig}
\usepackage{listings}
\usepackage{pgfplots, pgfplotstable}
\usepackage{diagbox} 
\usepackage[usestackEOL]{stackengine}
\usepackage{makecell}
\usepackage{mathrsfs}
\usepackage{moresize}
\usepackage{multirow}
\usepackage{multicol}
\usepackage[numbers]{natbib}
\usepackage[T1]{fontenc}
\usepackage{xcolor}
\allowdisplaybreaks[1]
\definecolor{orchid}{rgb}{0.7, 0.4, 1.1}
\DeclareCaptionFont{listing}{\footnotesize}
\DeclareCaptionFormat{listing}{\rule{0.47\textwidth}{0.5pt}\vskip#1#2#3}
\definecolor{comment_color}{rgb}{0, 0.5, 0}
\definecolor{keyword_color}{rgb}{0.3, 0, 0.6}
\definecolor{string_color}{rgb}{0.5, 0, 0.1}

\def\tsc#1{\csdef{#1}{\textsc{\lowercase{#1}}\xspace}}
\tsc{WGM}
\tsc{QE}

\begin{document}
\let\WriteBookmarks\relax
\def\floatpagepagefraction{1}
\def\textpagefraction{.001}





\shorttitle{A GPU-based Hydrodynamic Simulator with Boid Interactions}

\shortauthors{Xi Liu, Gizem Kayar, Ken Perlin}

\title{{\Large A GPU-based Hydrodynamic Simulator with Boid Interactions}}


\author[1]{\color{black}Xi Liu}
\author[2]{\color{black}Gizem Kayar}
\author[3]{\color{black}Ken Perlin}

\address{$^ a$ xl3504@nyu.edu, $^ b$ gk2409@nyu.edu, $^ c$ perlin@nyu.edu\\
New York University}

\begin{abstract}
We present a hydrodynamic simulation system using the GPU compute shaders of DirectX for simulating virtual agent behaviors and navigation inside a smoothed particle hydrodynamical (SPH) fluid environment with real-time water mesh surface reconstruction. The current SPH literature includes interactions between SPH and heterogeneous meshes but seldom involves interactions between SPH and virtual boid agents. The contribution of the system lies in the combination of the parallel smoothed particle hydrodynamics model with the distributed boid model of virtual agents to enable agents to interact with fluids. The agents based on the boid algorithm influence the motion of SPH fluid particles, and the forces from the SPH algorithm affect the movement of the boids. To enable realistic fluid rendering and simulation in a particle-based system, it is essential to construct a mesh from the particle attributes. Our system also contributes to the surface reconstruction aspect of the pipeline, in which we performed a set of experiments with the parallel marching cubes algorithm per frame for constructing the mesh from the fluid particles in a real-time compute and memory-intensive application, producing a wide range of triangle configurations. We also demonstrate that our system is versatile enough for reinforced robotic agents instead of boid agents to interact with the fluid environment for underwater navigation and remote control engineering purposes.
\end{abstract}



\maketitle


\section{Introduction}
\begin{figure}[h!]
    \centering
    \includegraphics[width = 0.49\textwidth, height = 0.3\textwidth]{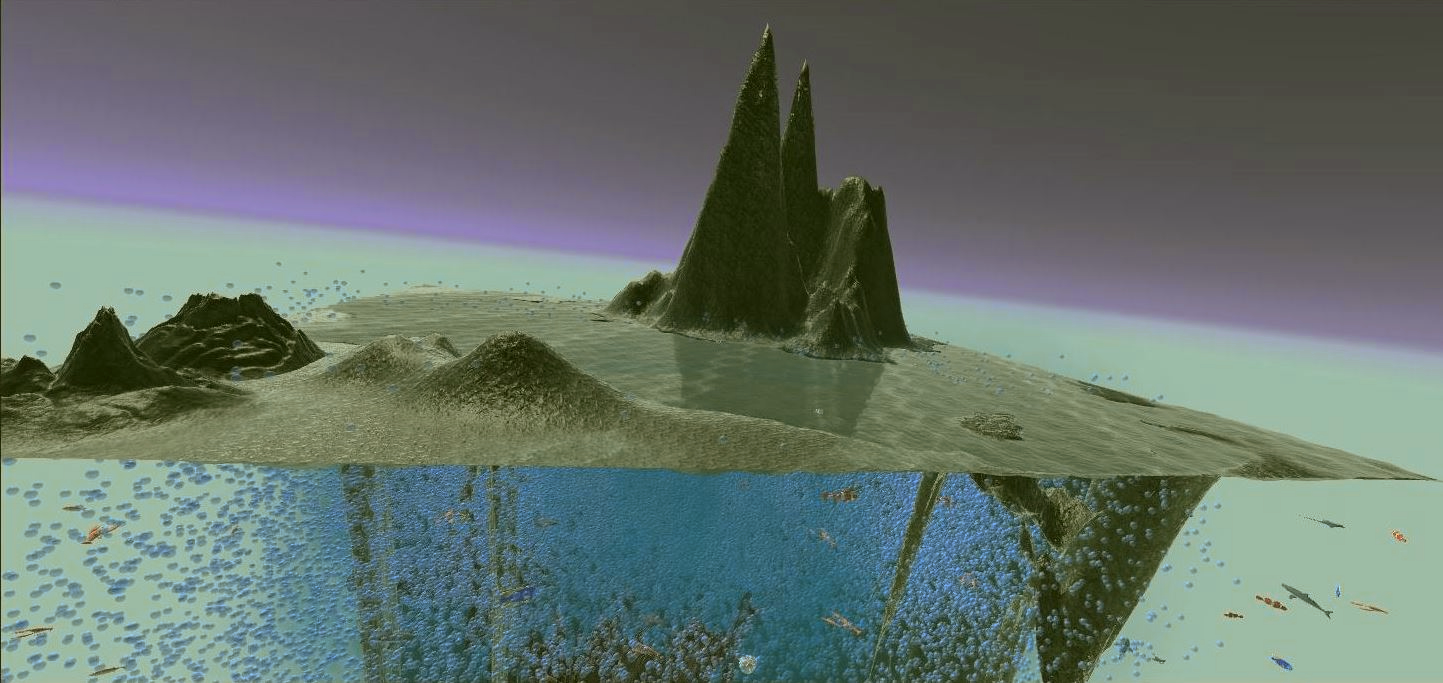}
    \caption{The simulation environment including the underwater fluid particles, terrain, marine creatures, etc.}
    \label{environment}
\end{figure}

\begin{figure}[h!]
    \centering
    \includegraphics[width = 0.49\textwidth, height = 0.3\textwidth]{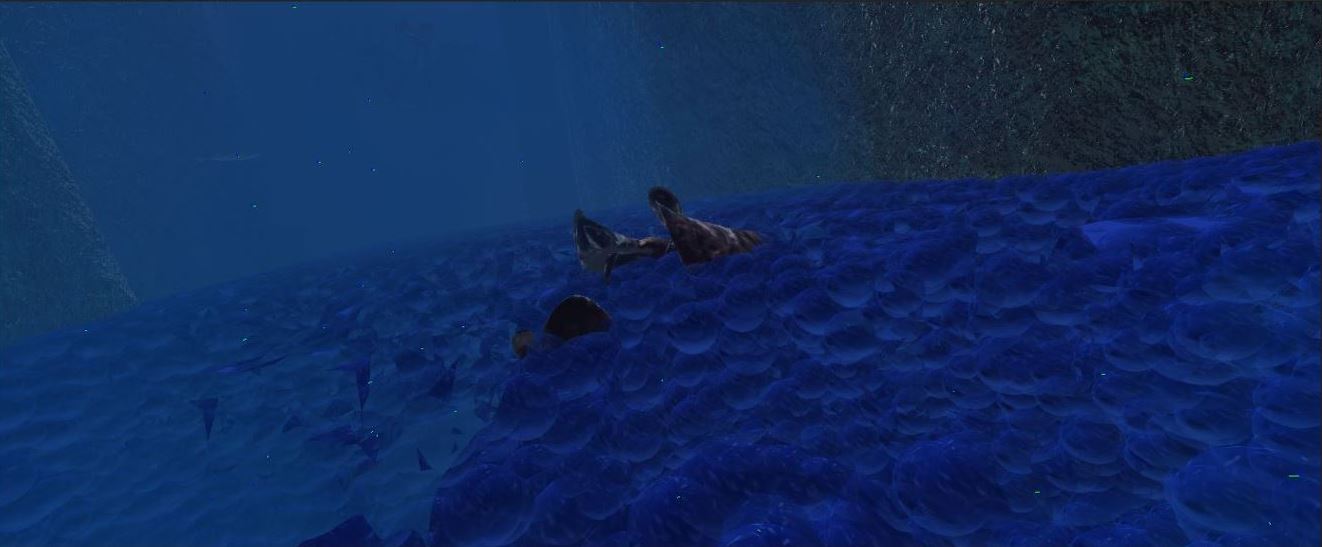}
    \caption{The underwater particle fluid substratum enclosed in a cavern-shaped terrain.}
    \label{substratum}
\end{figure}

\begin{figure}[h!]
    \centering
    \includegraphics[width = 0.49\textwidth, height = 0.3\textwidth]{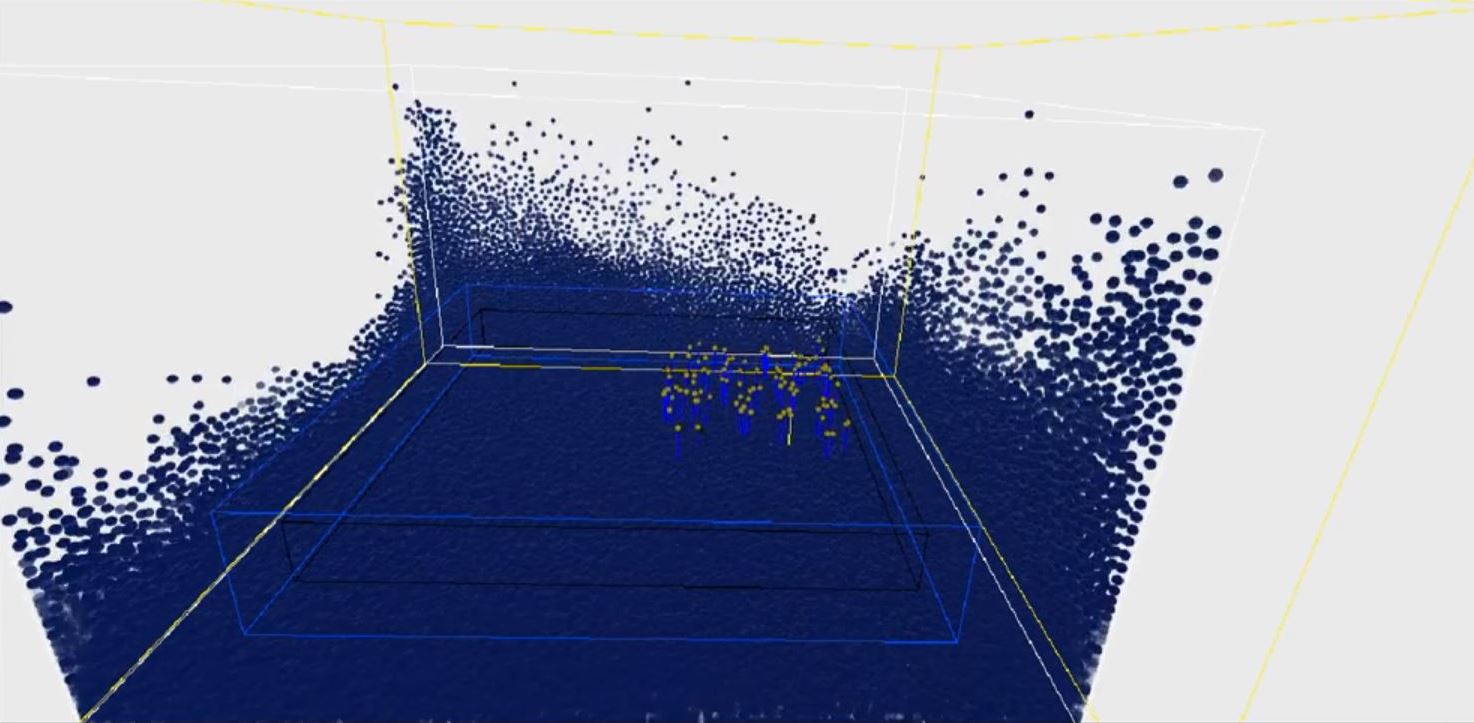}
    \caption{The SPH fluid particle and boid interactions in our system.}
    \label{the sph fluid particle}
\end{figure}

\begin{figure}
    \centering
    \includegraphics[width = 0.49\textwidth, height = 0.3\textwidth]{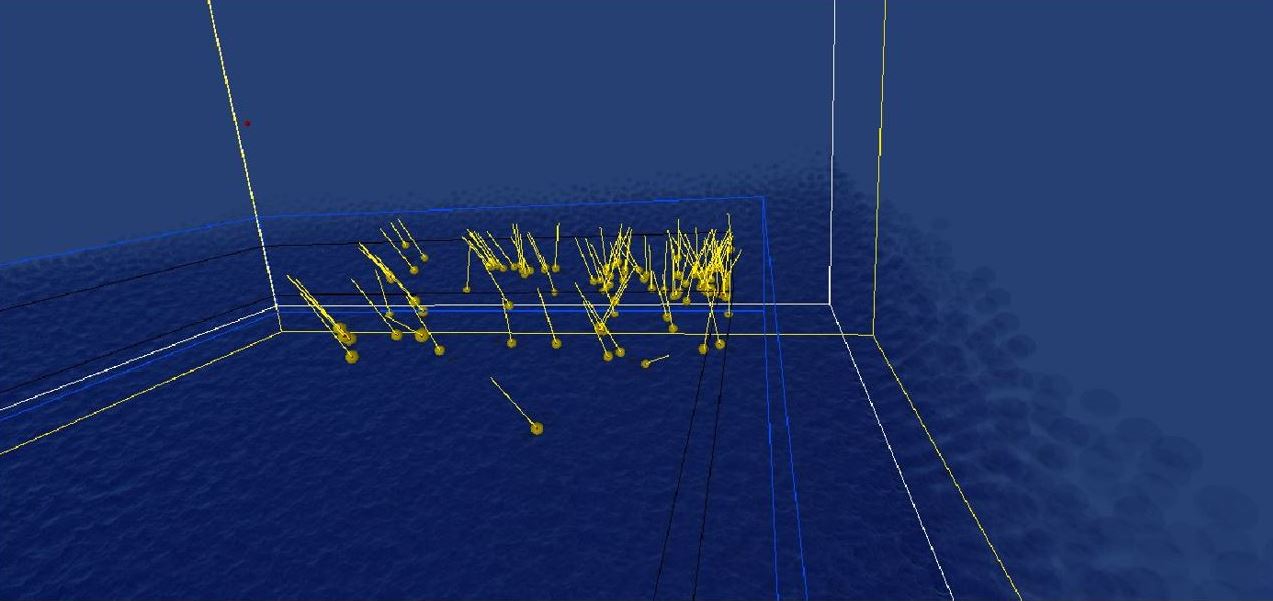}
    \caption{The blue objects are SPH particles. The yellow objects are boid agents.}
    \label{the blue objects}
\end{figure}

Parallel fluid mechanical models interacting with autonomous agents have played a crucial role in microfluidic biochemistry, atomic-level condensed matter quantum magnetohydrodynamics, and deep reinforcement learning. The modeling of self-propelled microbial agents undergoing quantum quorum sensing in bioorganic microfluidic environments involves bacterial agents swimming in massive amounts of molecules, in which the state variables of molecules need to be parallelized using molecular dynamical models. Remotely operated telerobotic agents that navigate in oceans on other celestial bodies, such as Titan or Enceladus, need to have a parallelized computational model of the fluid environment they will be operating in to make better autonomous decisions \cite{Davis_2023}, \cite{NASA_2015}. Here we consider the problem of computationally modeling the interactions between self-propelled or autonomous boid agents with fluid dynamical environments. Contemporary research in smoothed particle hydrodynamics (SPH) often explores interactions between fluid and heterogeneous meshes, but it rarely delves into the integration of SPH with virtual boid agents. In this context, our work stands out as it not only establishes a robust platform for examining the interactions between distributed virtual boid agents and diverse parallel fluid environments but also extends its applicability to simulate the behavior of agents within molecular-scale environments. By incorporating principles from molecular dynamics into the fluid simulator, we can uncover novel insights into the behaviors and interactions at the microscopic level and nanoscopic level, facilitating a deeper understanding of the underlying molecular mechanisms governing the behaviors of both viscoelastic medium and bioorganic agents. Our research builds a platform for testing the interaction between distributed virtual boid agents and different parallel fluid environments that can be used to simulate agents' self-driven or remote-controlled operations. The example usages are shown in figure \ref{environment}, \ref{substratum}, \ref{the sph fluid particle}, \ref{the blue objects}, etc. Our fluid simulation contains large datasets, such as three-dimensional grids and high-resolution meshes. GPUs offer a significant advantage in processing and manipulating such data due to their high memory bandwidth and ability to efficiently handle large arrays. This allows for faster data transfers and improved overall simulation performance.

In the simulation of the fluid environment, we are using particles to model the forces between the fluid elements, which is crucial to have parallelization. We decided to use GPU compute shaders of DirectX for this parallelization task. Along with tensor processing units and quantum processing units, graphics processing units utilize their special hardware design for computational acceleration, increasingly used more for tasks outside the rendering pipeline \cite{Aamodt_2018}. In our simulator, we assigned one GPU thread per particle in the neighborhood search of the SPH, boid, and marching cubes algorithms. In the integration of the parallel SPH model with the distributed boid model of virtual agents, the system allows for agent-fluid interactions, which involve a higher number of particles than using SPH for fluid particles alone, making parallelization more imperative. The boid-based agents influence the movement of SPH fluid particles, while the forces from the SPH algorithm affect the behavior of the boids, in which the interactions between the SPH particles and boid agents are also parallelized. We trained a rover to navigate towards an underwater target, actively engaging with parallel fluid particles in its path. It has learned to intelligently avoid obstacles by receiving a penalty for collisions, while being rewarded for proximity to the target located on the seafloor. The voxel density value of each cube corner is also computed in parallel, which is constructed from the signed distance field between the cube corner and the weighted average position of nearby neighbors. After obtaining the voxel density values from particle positions that define the fluid's location, we use the marching cubes algorithm to generate the triangles of the surface that wraps around the particles. The code of our system is located at https://github.com/xi-liu-cs/water.

\section{Background}
\emph{Lagrangian fluid} methods track information of individual particles including their position, velocity, pressure, and density as a function of time. Classical mechanics including conservation of mass and energy are directly applied per particle. The Lagrangian framework possesses difficulty in expression of stresses \cite{gingold-monaghan_sph}. \textit{Eulerian fluid} methods describe the flow velocity field as a function of space and time. Instead of tracking information per particle as in the Lagragian method, the Eulerian method tracks information such as velocity, acceleration of elements at a location of interest in a certain time \cite{foster_1996}. Quantum electrodynamical models including the incompressible Schrodinger flow \cite{Chern_2016} and Gross Pitaevskii equation \cite{Tsuzuki_2022} are also used in Eulerian and Lagrangian fluid simulations.

The \textit{smoothed particle hydrodynamics (SPH)} method makes use of the fluid elements of the Lagrangian description of flow. The movements of the fluid elements is modeled by Newtonian equations of forces from pressure gradient and body forces. Due to the complicated motion of large $n$-body systems, the method assumes that the locations of fluid parcels are randomly 
distributed based on density \cite{gingold-monaghan_sph} \cite{muller_2003}. High resolution SPH simulations involves high amount of particles, so parallel algorithms or machine learning methods are used for speedups.

Instead of using smoothed particle hydrodynamics (SPH) for fluid motion, we could use molecular dynamics (MD) models \cite{Gunsteren_1990}. SPH at the mesoscale deals with larger particles or fluid elements, which are more representative of macroscopic behaviors. MD at the microscale involves modeling individual molecules or atoms, capturing very fine-grained details. Although the SPH method has some difficulty in the imposition of boundary conditions, we apply an opposing damped force whenever a particle reaches a boundary according to Newton's third law. SPH simulations are computationally less demanding than MD simulations. This means that SPH can simulate larger systems and longer time scales more efficiently, making it suitable for studying phenomena that occur over extended periods or involve large volumes.

\textit{Machine learning fluid} methods can be used to predict particle attributes to reduce the large computational resources needed in addition to parallel computing speedups. A regression forest can be trained on video data obtained from the position based fluids \cite{Macklin_2013} algorithm to iteratively predict the particle states \cite{Ladicky_2015}. The smooth particle networks (SPNets) implement the position based fluids algorithm through a smooth particle convolution (ConvSP) layer that compute pairwise interactions between particles and a signed distance field convolution (ConvSDF) layer to compute interactions between particle and static objects, in which they used the Nvidia FleX \cite{macklin_2014} fluid simulator to generate the ground truth fluid parameters such as positions and velocities and uses backpropagation and gradient descent to update the fluid parameters iteratively until convergence \cite{Schenck_Fox_2018}. Long short term memory networks \cite{Wiewel_Becher_Thuerey_2019}, generative adversarial networks \cite{Xie_Franz_Chu_Thuerey_2018}, and neural style transfer \cite{Kim_Azevedo_Gross_Solenthaler_2020} can be used to perform inference and generate advected field quantities or particle attributes with temporal coherence.

Underwater agents and robotic behaviors in fluid simulations model the interaction and control of robots coupling with water. Starfish soft robot was used in a pipeline embedded with a differentiable simulator to alternate between simulated and real experiments, running system identification and trajectory optimization based on gradients from differentiable simulator in the simulation step, then let robotic hardware to execute the optimized trajectory underwater to collect new data as input to the simulation \cite{du_2021} \cite{du_2021_diffpd}. SPH can be used to control robotic swarms, in which robots are modeled as particles subject to physical forces, using two layers in the control architecture, the lower SPH layer controls the swarm behavior on a particle basis and upper layer control the macroscopic behavior of the swarm \cite{pac_2007}.

\textit{Boids} model the flock motion through group actions and interaction between individual birds' behaviors in a group to simulate the local perception of individual birds and aerodynamic flight, instead of scripting the path of individual agents \cite{reynolds_boids}. Autonomous robotic fish simulators simulate the swimming motion control and autonomous navigation under the influence of hydrodynamics models \cite{Liu_Hu_2004}. The boids algorithm implemented on the GPU in parallel can optimize the model’s performance, and a grid system can be used to optimize neighbor search \cite{reynolds_2006} and force fields can be utilized to allow for obstacle avoidance \cite{Chiara_Erra_Scarano_Tatafiore_1970}. The fluid structure interactions in fish swimming can be simulated with non-linear elastic solids with time varying field of distortions for fish body, and propulsive forces between the solid and fluid for the swimming motions \cite{curatolo_2016}.

Reynolds was one of the early pioneers in simulating flocking behavior, particularly in relation to birds \cite{reynolds_animation-with-scripts-and-actors, reynolds_boids}. The term "boids" refers to a generalized particle system where each boid represents a particle. Boids adhere to three main principles: collision avoidance (separation), velocity matching (alignment), and flock centering (cohesion). Reynolds described how each boid tracks its own position and velocity, updating its velocity based on these principles using user-defined hyperparameters  \cite{reynolds_steering}. Additional behaviors such as obstacle avoidance, fleeing, and pursuit can be achieved by incorporating simulated perception techniques like ray-casting. Viscek proposed an alternative approach that employs self-propelling particles and phase transitions \cite{vicsek_phase-self-driven-particles}. Modern approaches often combine or utilize variations of these methodologies. To optimize performance, one common technique is positional hashing, which reduces the time complexity of finding nearby neighbors. Reynolds initially used an iterative process to identify neighbors, resulting in quadratic complexity. Positional hashing involves mapping boids' approximate positions to grid cells, enabling the identification of neighbors within a subset of grid cells around a boid's current cell \cite{reynolds_crowds-ps3}. Other known methodologies for spatial hashing include quad/oct-trees \cite{shao-terzopoulos_autonomous-pedestrians} and navigation meshes \cite{kallmann_2014}.

\section{Governing Equations}
The forces acting on a moving fluid element consist of body forces and surface forces, which is modeled in the Navier Stokes equations. The body forces include gravitational, electric, and magnetic forces that act on the volumetric mass of the fluid element. The surface forces act on the surface of the fluid element due to pressure on the surface by outside surrounding fluid, and shear and normal stress distributions acting on the surface \cite{anderson_2010}. Let $\rho$ be the density, $\mathbf{V} = (u, v, w)$ be the flow velocity, $\frac{D}{Dt} = \frac{\partial}{\partial t} + \nabla\mathbf{V}$ be the substantial derivative, $p$ be the pressure, $\tau$ be the shear stress, $f$ be the body force per unit mass, from Newton's second law, we obtain
\begin{align}
\rho\frac{D\mathbf{V}}{Dt} &= -\nabla p + \nabla \cdot \tau + \rho f\\
\rho\frac{Du}{Dt} &= \frac{\partial(\rho u)}{\partial t} + \nabla\cdot(\rho u \mathbf{V})\\
&= \frac{\partial(\rho u)}{\partial t} + \frac{\partial(\rho u ^ 2)}{\partial x} + \frac{\partial(\rho u v)}{\partial y} + \frac{\partial(\rho u w)}{\partial z}\\
&= -\frac{\partial p}{\partial x} + \frac{\partial \tau_{xx}}{\partial x} + \frac{\partial \tau_{yx}}{\partial y} + \frac{\partial \tau_{zx}}{\partial z} + \rho f_x\\
\rho\frac{Dv}{Dt} &= \frac{\partial(\rho v)}{\partial t} + \nabla\cdot(\rho v \mathbf{V})\\
&= \frac{\partial(\rho v)}{\partial t} + \frac{\partial(\rho uv)}{\partial x} + \frac{\partial(\rho v ^ 2)}{\partial y} + \frac{\partial(\rho vw)}{\partial z}\\
&= -\frac{\partial p}{\partial y} + \frac{\partial \tau_{xy}}{\partial x} + \frac{\partial \tau_{yy}}{\partial y} + \frac{\partial \tau_{zy}}{\partial z} + \rho f_y\\
\rho\frac{Dw}{Dt} &= \frac{\partial(\rho w)}{\partial t} + \nabla\cdot(\rho w \mathbf{V})\\
&= \frac{\partial(\rho w)}{\partial t} + \frac{\partial(\rho uw)}{\partial x} + \frac{\partial(\rho vw)}{\partial y} + \frac{\partial(w ^ 2)}{\partial z}\\
&= -\frac{\partial p}{\partial z} + \frac{\partial \tau_{xz}}{\partial x} + \frac{\partial \tau_{yz}}{\partial y} + \frac{\partial \tau_{zz}}{\partial z} + \rho f_z
\end{align}
This code is an overview of our parallelized simulation process for particle-based fluid dynamics and agent interactions on the GPU. It employs parallel prefix sum, up sweep, and down sweep methods along with hashed cell indices for efficient neighbor list computation. The code then proceeds to calculate the contributions of neighboring particles and agents to various physical quantities, including density, pressure, and forces, while also considering interactions between boid agents, SPH fluid particles, and reinforcement learning forces. Finally, it updates the positions of particles, boid agents, and cube corners using numerical integration and computes the voxel density values and surface triangles using the marching cubes algorithm, enabling the generation of a detailed fluid mesh representation. Let $P$ be the set of particles, $B$ be the set of boid agents, $C$ be the set of cube corners of marching cubes.
\begin{lstlisting}[mathescape = true]
for particle$_i \in P$, boid$_j \in B$, corner$_k \in C$ of GPU thread$_{i, j, k}$ 
    find particle_neighbor_list$_i$, boid_neighbor_list$_j$, corner_neighbor_list$_k$ using the parallel prefix sum, up sweep, and down sweep method and parallel sorting of hashed cell indices
for particle$_i \in P$ of GPU thread$_i$
    for particle$_j \in$ neighbor_list$_i$
        compute the contribution of particle$_j$ to the density $rho_i$, pressure $p_i$ using the kernel
for particle$_i \in P$ of GPU thread$_i$
    for particle$_j \in$ neighbor_list$_i$
        compute the contribution of particle$_j$ to pressure forces, viscosity forces, external forces, and forces applied onto fluid particles by boid agents
for particle$_i \in P$, boid$_j \in B$, corner$_k \in C$ of GPU thread$_{i, j, k}$
    update position of particle$_i$, boid$_j$, corner$_k$ using numerical integration
for boid$_i$ in boid_agents of GPU thread$_i$
    for boid$_j \in$ boid_neighbor_list$_j$
        compute contributing forces based on interaction with SPH fluid particles, boid-boid interactions, and reinforcement learning forces
for cube_corner$_i \in C$ of GPU thread$_i$
    for particle$_j \in$ corner_neighbor_list$_i$
        compute contribution to the voxel density value based on the signed distance field between cube_corner$_i$ and particle$_j$
for cube_corner$_i \in C$ of GPU thread$_i$
    use marching cubes algorithm to generate surface triangles that encapsulate the fluid particles
\end{lstlisting}

\section{Smoothed Particle Hydrodynamics}
Let $A(\mathbf{x})$ be a continuous compactly supported function. The convolution between two functions $f$ and $g$ is defined as $(f * g)(t) = \int_{-\infty} ^ {\infty} f(\tau)g(t - \tau)d\tau$. Let $\varepsilon \in \mathbb{R} ^ +$, the convolution of $A(\mathbf{x})$ with the Dirac $\delta$ distribution is equal to $A$ itself:
\begin{align}
(A * \delta)(\mathbf{r}) &= \int A(\mathbf{r}')\delta(\mathbf{r - r'})d\mathbf{r}'\\
&= \lim_{\varepsilon \rightarrow 0} \int_{-\infty} ^ {r - \varepsilon} A(\mathbf{r'})\delta(\mathbf{r - r'})d\mathbf{r}'\\
&\quad + \int_{r - \varepsilon} ^ {r + \varepsilon} A(\mathbf{r}')\delta(\mathbf{r - r'})d\mathbf{r}'\\
&\quad + \int_{r + \varepsilon} ^ {\infty} A(\mathbf{r'})\delta(\mathbf{r - r'})d\mathbf{r}'\\
&= \lim_{\varepsilon \rightarrow 0} \int_{r - \varepsilon} ^ {r + \varepsilon} A(\mathbf{r}')\delta(\mathbf{r - r'})d\mathbf{r}'\\
&= A(\mathbf{r})\int \delta(\mathbf{r} - \mathbf{r}')d\mathbf{r}'\\
&= A(\mathbf{r})
\end{align}

Let $d \in \mathbb{Z}, W: \mathbb{R} ^ d \times \mathbb{R} ^ + \rightarrow \mathbb{R}$ is kernel function or smoothing kernel
{\small
\begin{align}
A(\mathbf{r}) \approx (A * W)(\mathbf{r}) &= \int A(\mathbf{r}')W(\mathbf{r - r'}, h)d\mathbf{r}'
\end{align}
}
We typically use a distribution that is close to a Gaussian distribution to approximate the Dirac $\delta$ distribution. Let $W$ be a function satisfying $\int W(\mathbf{r})d\mathbf{r} = 1$, $\mathbf{r}$ be the center of mass, the smoothed density $\rho_s(\mathbf{r})$ is
{\small
\begin{align}
\rho_s(\mathbf{r}) &= \int W(\mathbf{r} - \mathbf{r}')\rho(\mathbf{r}')d\mathbf{r}'
\end{align}
}
The fluid quantities are interpolated and spatial derivatives are approximated with finite number of sample positions of adjacent particles. Let $\mathcal{F}$ be the set containing all point samples, $v_i$ be the volume of particle $i$ with $\rho_i = m_i / v_i$ \cite{müller_2003}, the analytic integral is approximated by a sum over sampling points:
\begin{align}
A(\mathbf{r}) &= (A * W)(\mathbf{r}_i)\\
&= \int A(\mathbf{r}')W(\mathbf{r}_i - \mathbf{r}', h)d\mathbf{r}'\\
&\approx \sum_{j \in \mathcal{F}}A_j W(\mathbf{r}_i - \mathbf{r}_j, h) v_j\\
&= \sum_{j \in \mathcal{F}}A_j W(\mathbf{r}_i - \mathbf{r}_j, h)\frac{m_j}{m_j / v_j}\\
&= \sum_{j \in \mathcal{F}}A_j W(\mathbf{r}_i - \mathbf{r}_j, h)\frac{m_j}{\rho_j}\\
&= \sum_{j \in \mathcal{F}}A_j \frac{m_j}{\rho_j}W(\mathbf{r}_i - \mathbf{r}_j, h)
\end{align}

All field quantities indexed using a subscript denote the field evaluated at the respective position. $A_j = A(\mathbf{r}_j), W_{ij} = W(\mathbf{r_i - r_j}, h)$. Each particle $j$ contain information of its location $\mathbf{r}_j$, mass $m_j$, and field $A_j$. It is optional to storing the density $\rho_j$ since it can be reconstructed from location and mass.

The kernel $W$ are chosen such that it approximates the Dirac Delta function. The choice of the kernel in this paper is based on a combination of Matthias Muller's work and experimental work \cite{müller_2003}. The kernel $W$ in a SPH simulation can be represented by the form below, \cite{bindel_2011}. Let $d \in \mathbb{Z}$ be the dimension, $f: \mathbb{R} ^ d \times \mathbb{R} ^ + \rightarrow 
\mathbb{R}$ be a function, $h$ is the kernel's smoothing length, $r$ is the center of mass, $c$ be a scaling factor
\begin{align}
W(r, h) &= \frac{1}{ch ^ d}\begin{cases}
f(r, h) & 0 \leq r \leq h\\
0 & \text{otherwise}
\end{cases}\\
W_{\text{poly6}}(r, h) &= \frac{315}{64\pi h ^ 9}
\begin{cases}
(h ^ 2 - r ^ 2) ^ 3 & 0 \leq r \leq h\\
0 & \text{otherwise}\\
\end{cases}\\
W_{\text{cubic}}(r, h) &= \begin{cases}
\frac{2}{3} - r ^ 2 + \frac{1}{2}r ^ 3 & 0 \leq r \leq 1\\
\frac{1}{6}(2 - r) ^ 3 & 1 \leq r \leq 2\\
0 & r > 2
\end{cases}
\end{align}

\section{Pressure and Viscosity}
We can substitute the pressure $p$ term for the continuous compactly supported function $A$ in the equation $A(\mathbf{r}) \approx \sum_{j \in \mathcal{F}}A_j \frac{m_j}{\rho_j}W(\mathbf{r}_i - \mathbf{r}_j, h)$ that is derived earlier to obtain
\begin{align}
f_i ^ {\text{pressure}} &= -\nabla p(\mathbf{r}_i) = -\sum_j m_j \frac{p_j}{\rho_j}\nabla W(\mathbf{r}_i - \mathbf{r}_j, h)
\end{align}

This force is asymmetric when only two particles interact \cite{müller_2003} \cite{bindel_2011}. One way to symmetrize it is taking the arithmetic mean
\begin{align}
f_i ^ {\text{pressure}} &= -\sum_j m_j \frac{p_i + p_j}{2\rho_j}\nabla W(\mathbf{r}_i - \mathbf{r}_j, h)
\end{align}

The ideal gas equation of state $p_i = k(\rho_i - \rho_0)$ is used to compute the pressure, where $k$ is a temperature dependent gas constant, $\rho_0$ is the rest density \cite{desbrun_1996}.
\begin{align}
W(\mathbf{r}, h) &= \frac{1}{ch ^ d}
\begin{cases}
f(q) & 0 \leq q \leq 1\\
0 & \text{otherwise}
\end{cases}\\
q &= \|\mathbf{r}\| / h \in \mathbb{R}, d \in \mathbb{Z}\\
\nabla W_{spiky}(\mathbf{r}, h) &= -\frac{30}{\pi h ^ 4}\frac{(1 - q) ^ 2}{q}\mathbf{r}\\
f_i ^ {\text{pressure}} &= -\sum_j m_j \frac{p_i + p_j}{2\rho_j}\nabla W_{spiky}(\mathbf{r}_i - \mathbf{r}_j, h)
\end{align}
substitute the ideal gas equation for $p_i$ and $p_j$, and expand the $\nabla W_{spiky}$ term
{\small\begin{align}
f_i ^ {\text{pressure}} &= \sum_j m_j \frac{k(\rho_i - \rho_0) + k(\rho_j - \rho_0)}{2\rho_j}\left(\frac{30}{\pi h ^ 4}\frac{(1 - q_{ij}) ^ 2}{q}\mathbf{r_{ij}}\right)\\
&= \frac{15k}{\pi h ^ 4}\sum_{j \in N_i} m_j \frac{\rho_i + \rho_j - 2\rho_0}{\rho_j}\frac{(1 - q_{ij}) ^ 2}{q_{ij}}\mathbf{r}_{ij}
\end{align}}

where $q_{ij} = \|\mathbf{r}_{ij}\| / h, \mathbf{r}_{ij} = \mathbf{r}_i - \mathbf{r}_j$, $N_i$ is the neighbor list of particle $i$ within the smoothing length $h$

We can substitute the viscosity term $\mu \nabla ^ 2 v$ for the continuous compactly supported function $A$ in $A(\mathbf{r}) \approx \sum_{j \in \mathcal{F}}A_j \frac{m_j}{\rho_j}W(\mathbf{r}_i - \mathbf{r}_j, h)$ that is derived earlier to obtain
\begin{align}
f_i ^ {\text{viscosity}} &= \mu \nabla ^ 2 v(\mathbf{r}_a) = \mu \sum_j m_j \frac{v_j}{\rho_j} \nabla ^ 2 W(\mathbf{r}_i - \mathbf{r}_j, h)
\end{align}

Viscosity forces are dependent on velocity differences between particles, so can symmetrize the viscosity forces using velocity differences
\begin{align}
f_i ^ {\text{viscosity}} &= \mu \sum_j m_j \frac{v_i - v_j}{\rho_j} \nabla ^ 2 W(\mathbf{r}_i - \mathbf{r}_j, h)\\
\nabla ^ 2 W_{viscosity} &= \frac{40}{\pi h ^ 4} (1 - q)\\
f_i ^ {\text{viscosity}} &= \mu \sum_j m_j \frac{v_i - v_j}{\rho_j} \left(\frac{40}{\pi h ^ 4} (1 - q_{ij})\right)\\
&= \frac{40\mu}{\pi h ^ 4}\sum_{j \in N_i}  m_j \frac{v_i - v_j}{\rho_j}(1 - q_{ij})
\end{align}
\begin{align}
&f_i ^ {\text{pressure}} + f_i ^ {\text{viscosity}}\\
&= \frac{15k}{\pi h ^ 4}\sum_{j \in N_i} m_j \frac{\rho_i + \rho_j - 2\rho_0}{\rho_j}\frac{(1 - q_{ij}) ^ 2}{q_{ij}}\mathbf{r}_{ij}\nonumber\\
&\quad + \frac{40\mu}{\pi h ^ 4}\sum_{j \in N_i}  m_j \frac{v_i - v_j}{\rho_j}(1 - q_{ij})\\
&= \frac{m_j(1 - q_{ij})}{\pi h ^ 4 \rho_j} \cdot\\
&\qquad\left(\frac{15k(\rho_i + \rho_j - 2\rho_0)(1 - q_{ij})\mathbf{r}_{ij}}{q_{ij}} - 40\mu v_{ij}\right)
\end{align}
\section{Underwater Agent Behaviors}
At the start of the simulation, \(n\_boid \in \mathbb{N}\) boids are initialized in randomized positions around a 3D volume, denoted as the \emph{Boid Space}. The \emph{Boid Space} is discretized into \(g_x \times g_y \times g_z\) grid cells, in which $g_x, g_y, g_z$ are number of cells in each dimension. These metrics are user-defined hyper-parameters that can be adjusted to suit the situation. Spatial hashing of any 3D vector position \(p_{x,y,z}\) within the \emph{Boid Space} into \(p_{hx,hy,hz}\) is defined as follows:
\begin{align}
p_{hx,hy,hz}(p_{x,y,z}) &=
\begin{cases}
\lfloor \frac{p_x}{x} \cdot g_x \rfloor\\
\lfloor \frac{p_y}{y} \cdot g_y \rfloor\\
\lfloor \frac{p_z}{z} \cdot g_z \rfloor
\end{cases}
\end{align}

The \emph{Boid Grid} tracks the number of boids within each grid cell. If obstacles with meshes are present in the simulation, each obstacle's mesh helps to define a separate \emph{Obstacle Grid}. The mesh normals of each obstacle are attributed to grid cells in the \emph{Obstacle Grid} and contribute to obstacle avoidance in boids that are close to or within the same grid cell as these mesh normals.
\begin{figure}[h!]
    \centering
    \includegraphics[width = 0.49\textwidth, height = 0.3\textwidth]{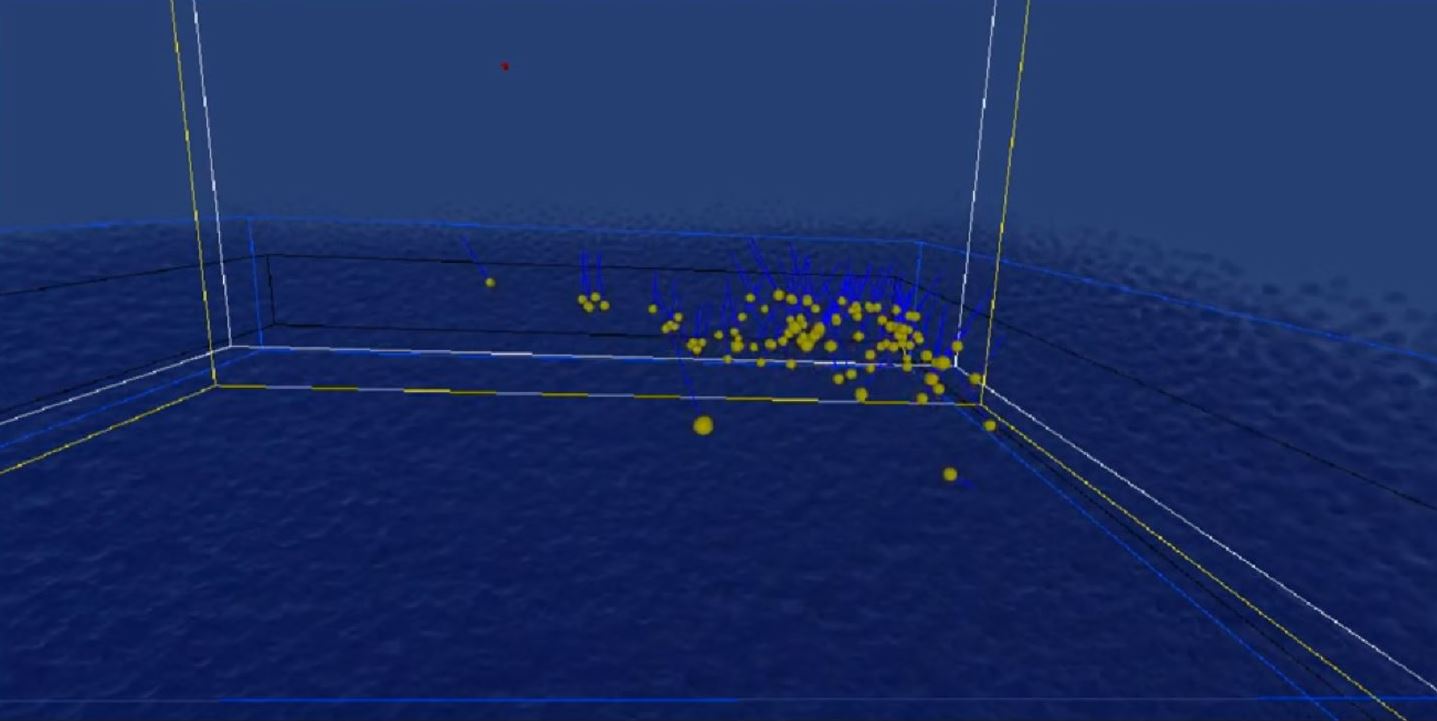}
    \caption{SPH boid interactions}
    \label{sph boid interactions}
\end{figure}
\begin{figure}
    \centering
    \includegraphics[width = 0.49\textwidth, height = 0.3\textwidth]{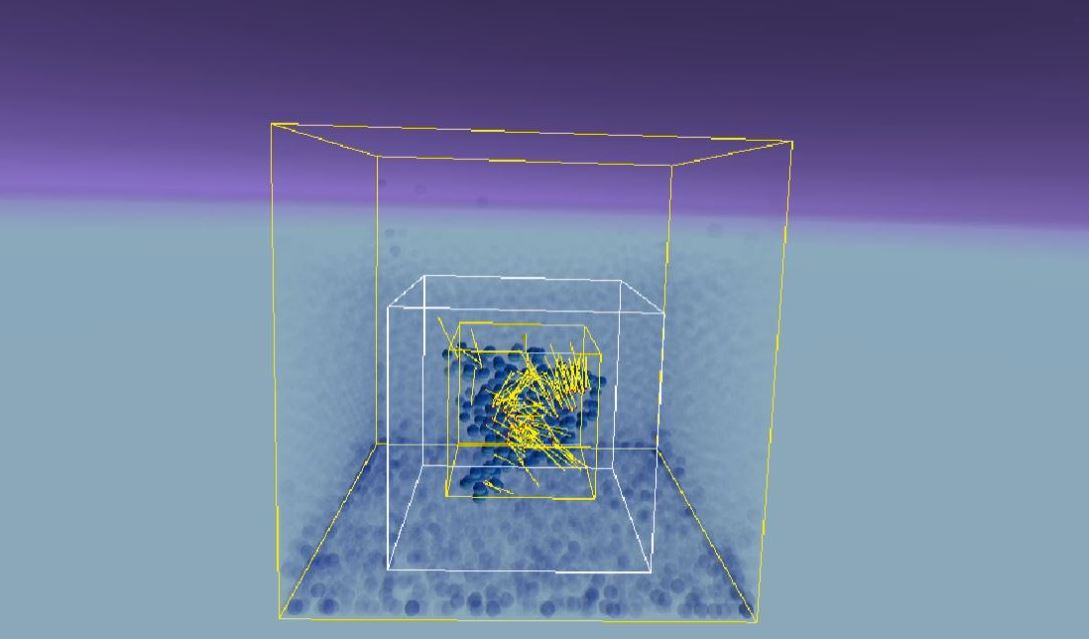}
    \caption{A SPH particle has higher opacity if it is touched by a boid.}
    \label{a sph particle has higher opacity}
\end{figure}
\section{Boid Motion and Agent Fluid Interaction}
Boids have a "visual range" that represents the distance at which the other boids are considered "neighbors". Within the visual range there is a smaller "protection range" wherein boids inside the protection range are considered too close. Neighboring boids in the protection range influence the boid's \emph{Separation} behavior; neighbors outside the protection range influence the boid's \emph{Alignment} and \emph{Cohesion} behavior. Obstacles are also "tracked" by each boid as long as they are within the Boid's visual range and influence the boid's \emph{Obstacle Avoidance} behavior. Such behaviors are weighted by hyper-parameters tuned by the user. In a hashed spatial grid setup, the visual and protection ranges can be represented by some set number of grid cells extending outward from the boid's current grid cell. This range is also defined as a hyper-parameter, adjustable by any user. In our system, the update loop of a boid include these steps:
\begin{enumerate}
    \item \emph{Separation}: The boid adjusts its velocity $v$ to move away from any neighbor boids in the protection range based on the position $p$ differences between the boids within a neighborhood, $sep$ is an avoid separation scaler that is a tunable paramter, and $protect\_range$ is the threshold distance value within which the boids need to fly away from each other.
\begin{lstlisting}
for all boid_i in parallel do
    close = vec3(0)
    sep = 0
    for boid_j $\in$ boid_i.neighbor_list
        d = boid_i.p - boid_j.p
        if(length(d) < protect_range)
            close += d
            sep = clamp(1 - d / protect_range, 0, 1) 
    boid_i.v += sep * close
\end{lstlisting}
    \item \emph{Alignment}: The boid adjusts its velocity to match the motion direction of all boids within its visual range. Let $v_a$ be the average velocity of boids inside the visible range of a particular boid. The $visible\_range$ parameter is the threshold distance value within which each boid attempts to match the velocity of other boids, and $mat$ is a matching factor that is a tunable parameter.
\begin{lstlisting}
for all boid_i in parallel do
    v_a = vec3(0)
    count = 0
    for boid_j $\in$ boid_i.neighbor_list
        d = boid_i.p - boid_j.p
        if(length(d) < visible_range)
            v_a += boid_j.v
            ++count
    if(count > 0) v_a /= count
    boid_i.v += mat * (v_a - boid_i.v)
\end{lstlisting}
    \item \emph{Cohesion}: The boid adjusts its velocity to move closely to the center of mass among itself and all other boids in the visual range. Let $p_a$ be the average position of boids inside the visible range of a particular boid, and $cen$ is the centering factor that is a tunable parameter.
\begin{lstlisting}
for all boid_i in parallel do
    p_a = vec3(0)
    count = 0
    for boid_j $\in$ boid_i.neighbor_list
        d = boid_i.p - boid_j.p
        if(length(d) < visible_range)
            p_a += boid_j.p
            ++count
    if(count > 0) p_a /= count
    boid_i.v += cen * (p_a - boid_i.p)
\end{lstlisting}
    \item \emph{Edge Avoidance}: The boid adjusts its velocity to return into the bounds of the \emph{Boid Space} if it is outside such bounds.
    \item \emph{Speed Limit}: The boid's velocity is clamped within a minimum and maximum speed.
    \item \emph{Position update}: The boid's position is updated based on the boid's updated velocity.
\end{enumerate}

\begin{itemize}
    \item \emph{Obstacle Avoidance}: The boid adjusts its velocity based on the averaged norm vector of each obstacle grid cell within the boid's visual range. Obstacles in the boid's protection range are weighed heavier than those in just the visual range.
    \item \emph{Hash Position Update}: The boid updates its hash position and informs the \emph{Boid Grid} if its hash position has shifted to a new grid cell.
\end{itemize}

The interaction between the fluid particles and boids involve several aspects: the first aspect is the mechanism for boid motions to influence the fluid particle motions; the second aspect is the mechanism for fluid particle motions to influence the boid motions. See figure \ref{sph boid interactions}, \ref{a sph particle has higher opacity}. To enable the interaction, the particle struct is defined as the following. The $touched\_by\_boid$ is an integer counter of the number of boids that are currently inside the neighborhood of a given fluid particle. $boid\_influence$ is the cumulative acceleration accounting for the force that each boid in the neighbor list applies to the fluid particle. The $cell\_index$ variable represents the cell that the particle belongs to, and $index\_in\_cell$ represents the index inside the cell that particle the is locating at, $neighbor$ is the count of number of neighbors that is currently found in the neighbor search, in which the neighbor search algorithm is covered in detail in the next section.

\begin{lstlisting}[caption = Particle structure]
struct particle
    float mass, density, pressure, speed;
    float3 position, velocity, acceleration, boid_influence;
    int neighbor, cell_index, index_in_cell, is_boid, touched_by_boid;
\end{lstlisting}

We have a GPU kernel $initialize\_particle$ that initialize the particle positions with a float3 vector with random numbers generated within the bounding box's boundary, the random number generation is using the exclusive or shift algorithm \cite{Reed_2013}. To implement the forces that the boids are applying to the fluid particles, we attach a particle to each boid using $particles[id.x].position = boids[id.x].position$ inside the Euler Cromer integration kernel dispatch. We use an integer $is\_boid$, in which if it is one, this particle is attached to a boid, if it is zero, then this particle is not attached to a boid. During the $initialize\_particle$ kernel, we allocate a portion of the $particles$ array to be specifically used as boid attachments through $if(id.x < n\_boid) particles[id.x].is\_boid = 1$. After attaching particles to boids, the motions of the boids from the boid algorithm affects the motion of fluid particles from the SPH algorithm.

\begin{lstlisting}[caption = Initialize particle]
uint s, seed;
float ran()
    s ^= 2747636419u; s *= 2654435769u; s ^= s >> 16;
    s *= 2654435769u; s ^= s >> 16; s *= 2654435769u;
    return s / 4294967295.0;

float r(float a, float b) return a + ran() * (b - a);

#pragma kernel initialize_particle
[numthreads(thread_per_group, 1, 1)]
void initialize_particle(uint3 id : SV_DispatchThreadID)
    if(id.x >= n_particle) return;
    s = seed + id.x;
    let $\varepsilon > 0$;
    particles[id.x].position = $(r(x_{min} + \varepsilon, x_{max} - \varepsilon)$, $r(y_{min} + \varepsilon, y_{max} - \varepsilon)$, $r(z_{min} + \varepsilon, z_{max} - \varepsilon));$
    if(id.x < n_boid) particles[id.x].is_boid = 1;
    else particles[id.x].is_boid = 0;
    particles[id.x].touched_by_boid = 0;
    particles[id.x].boid_influence = float3(0., 0., 0.);
\end{lstlisting}

To compute the acceleration, we use the equation for $f_i ^ {pressure} + f_i ^ {viscosity}$ derived in the earlier sections of pressure and viscosity. To account for the forces applied by the boids onto the particles, we increment the $touched\_by\_boid$ counter whenever we found another boid inside the neighbor list of the given particle, and accumulate the $boid\_influence$ acceleration, which is then integrated in the Euler Cromer kernel dispatch.
\begin{lstlisting}[caption = Compute acceleration]
float3 acceleration(int a, int b)
    float rho_i = particles[a].density,
    rho_j = particles[b].density;
    float3 diff_pos = particles[a].position - particles[b].position,
    diff_vel = particles[a].velocity - particles[b].velocity;
    float diff_square = dot(diff_pos, diff_pos);
    if(diff_square > radius2 || diff_square == 0) return float3(0., 0., 0.);
    float q = sqrt(diff_square) / radius,
    q2 = 1 - q;
    float3 res = mass * q2 / (pi * radius4 * rho_j)
    * (15 * bulk_modulus * (rho_i + rho_j - 2 * rest_density)
    * q2 * diff_pos / q - 40 * viscosity_coefficient * diff_vel);
    if(particles[b].is_boid > 0)
        ++particles[a].touched_by_boid;
        particles[a].boid_influence += res;
    return res;
\end{lstlisting}

\begin{figure}[h!]
    \centering
    \includegraphics[width = 0.49\textwidth, height = 0.3\textwidth]{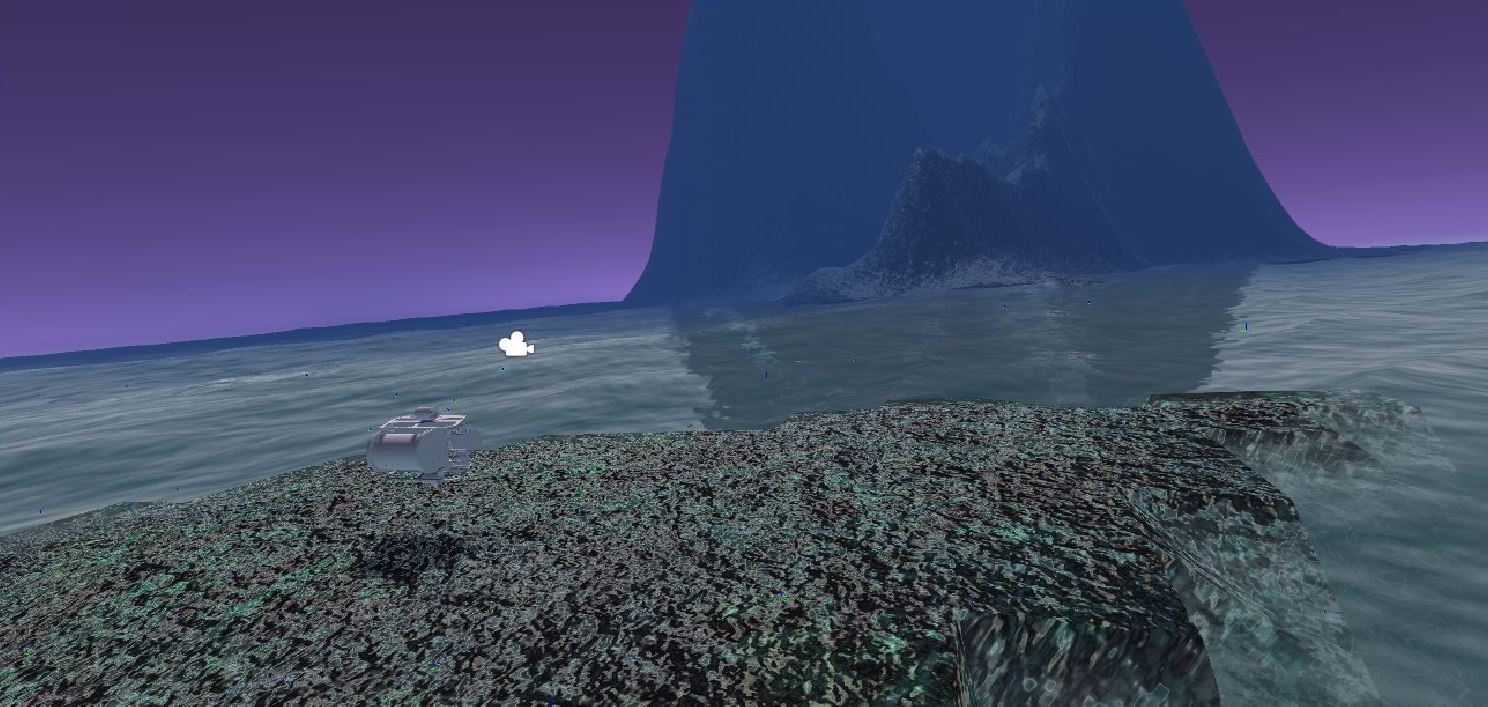}
    \caption{The rover's initial location is on an island.}
    \label{the rover's initial location}
\end{figure}

To enable SPH fluid particles to affect the boids, we modify the velocity of the boids to account for the influence of each SPH fluid particle. We used a variable $sph\_factor$ which is a tunable factor to control the degree of influence that the SPH fluid particles are affecting the boids, then we used $boids[id.x].velocity += particles[id.x].velocity * sph\_factor$ to implement this aspect of the interaction. To enable SPH fluid particles to interact with the rover, we apply the same algorithm in which we attach particles to each rover to enable each rover to affect the SPH fluid particles, and we modify the velocity of the rover based on the velocity of SPH fluid particles. We use the robot operating system \cite{Quigley_2007} and Unity machine learning agents library \cite{Juliani_2020} (built-in functions SetReward, EndEpisode, AddObservation, etc) to control the rover motion. The rover have its own camera, separate from the main camera of the simulation. The rover's initial position is on an island. See figure \ref{the rover's initial location}. We trained the rover to reach an underwater target in which it interacts with the underwater fluid particles and it is trained to find a path that avoids obstacles by applying a penalty whenever the rover collides with an obstacle and applying a reward whenever the agent is close enough to the target at the seafloor, in which we detected the collision through the $is\_overlap$ function defined below. Let $S$ be a set of states, $A$ be a set of actions, $\theta$ be the parameters of a policy $\pi$, $T \in \mathbb{N}$ be the number of time steps, $\gamma \in [0, 1)$ be the discount factor. The rover attempts to learn a policy $\pi_{\theta}: S \times A \rightarrow \mathbb{R}$ that maximizes the cumulative discounted reward $\sum_{t = 0} ^ {T - 1} \gamma ^ t R(S_t, A_t)$. The rover and the fluid environment engage in a series of discrete time steps, represented as $t \in \mathbb{N}$. At each time step $t$, the rover observes the state of the environment, denoted as $S_t \in S$, and based on this observation, selects an action $A_t \in A(S)$. After one time step, as a consequence of its action, the rover receives a numerical reward $R_{t + 1} \in \mathbb{R}$ and transitions to a new state $S_{t + 1} \in S$.

\begin{lstlisting}[caption = Rover behavior]
int overlap = 0;
public bool is_overlap() return overlap > 0;
void OnTriggerEnter(Collider other) ++overlap;
void OnTriggerExit(Collider other)
    --overlap;
    if(String.Equals(other.gameObject.name, "Landscape"))
        SetReward(-1.0f);
        EndEpisode();

public override void OnEpisodeBegin()
    transform.position = initial_position;

public override void CollectObservations(VectorSensor sensor)
    sensor.AddObservation(target.localPosition);
    sensor.AddObservation(this.transform.localPosition);
    sensor.AddObservation(rigid_body.velocity);

public override void OnActionReceived(ActionBuffers actions)
    float dx = actions.ContinuousActions[0],
    dy = actions.ContinuousActions[1],
    dz = actions.ContinuousActions[2];
    transform.position += new Vector3(dx, dy, dz) * Time.deltaTime * speed;

    float distance_to_target = Vector3.Distance(this.transform.localPosition, target.localPosition);
    if(distance_to_target < 1f)
        SetReward(1f);
        EndEpisode();
    if(is_overlap() || is_outside(this.transform.localPosition))
        SetReward(-1f);
        EndEpisode();
\end{lstlisting}
\section{Neighborhood Search}
\begin{figure}[h!]
    \centering
    \includegraphics[width = 0.23\textwidth, height = 0.15\textwidth]{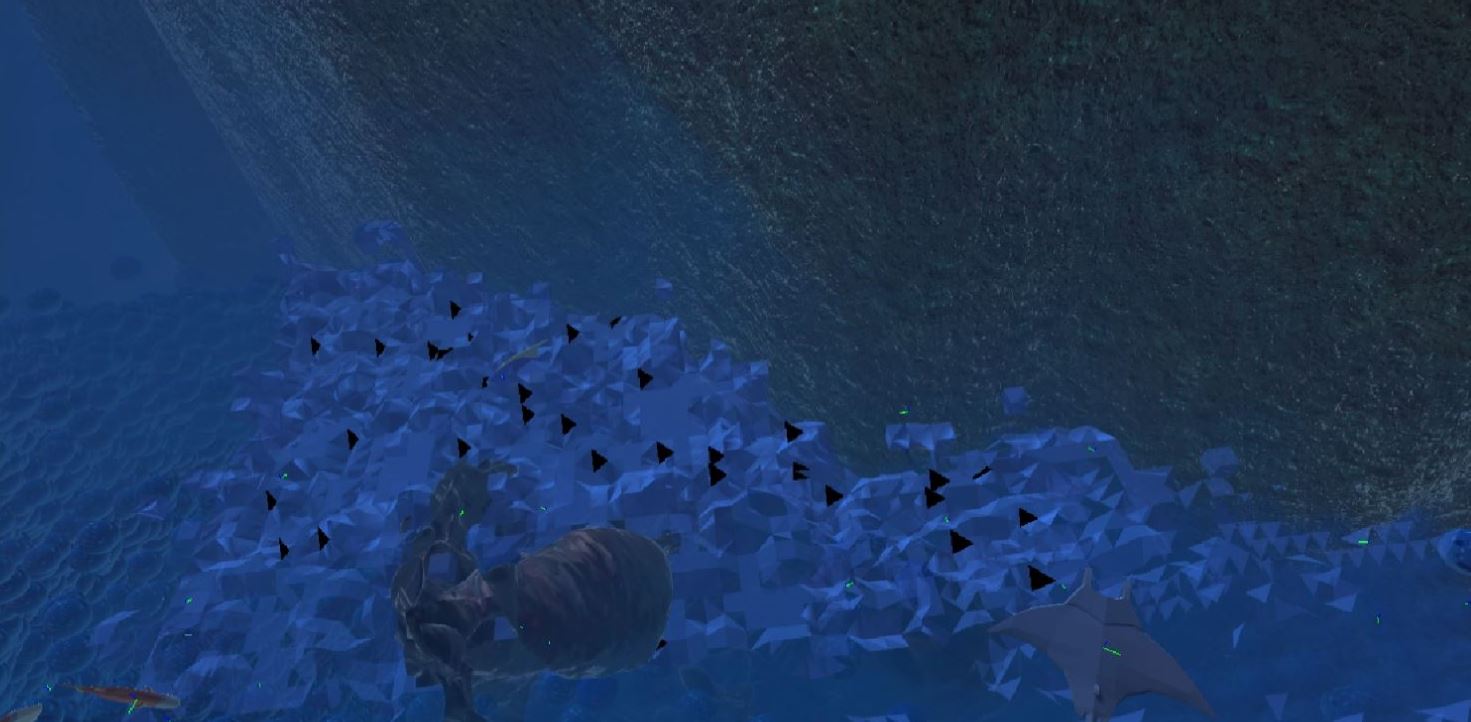}
    \includegraphics[width = 0.23\textwidth, height = 0.15\textwidth]{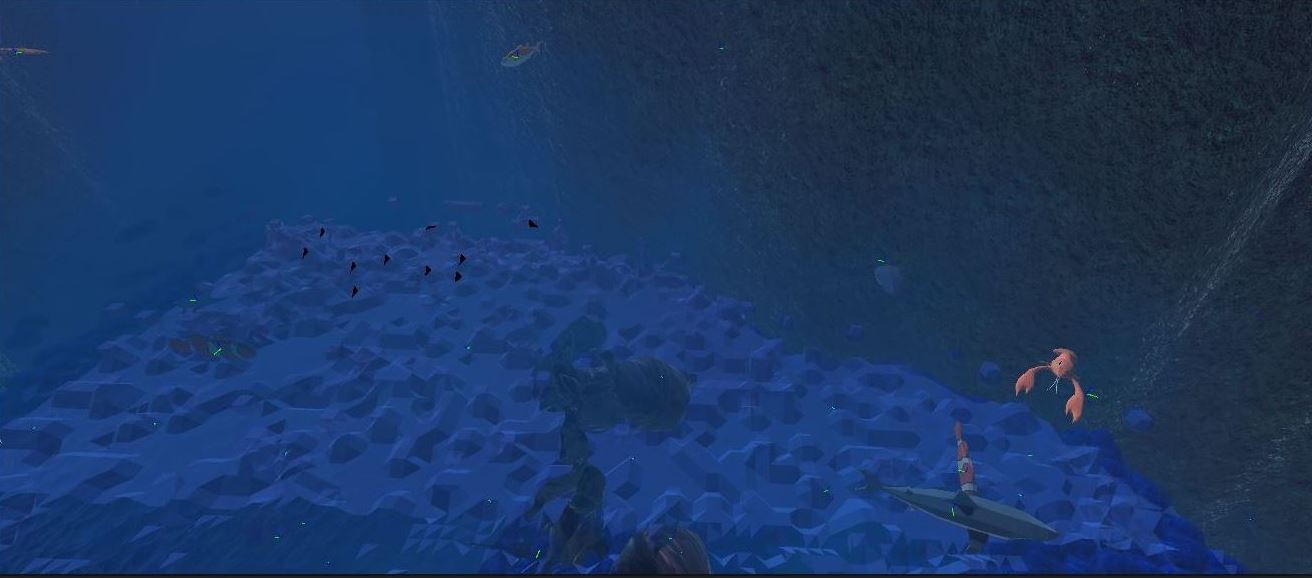}
    \includegraphics[width = 0.23\textwidth, height = 0.15\textwidth]{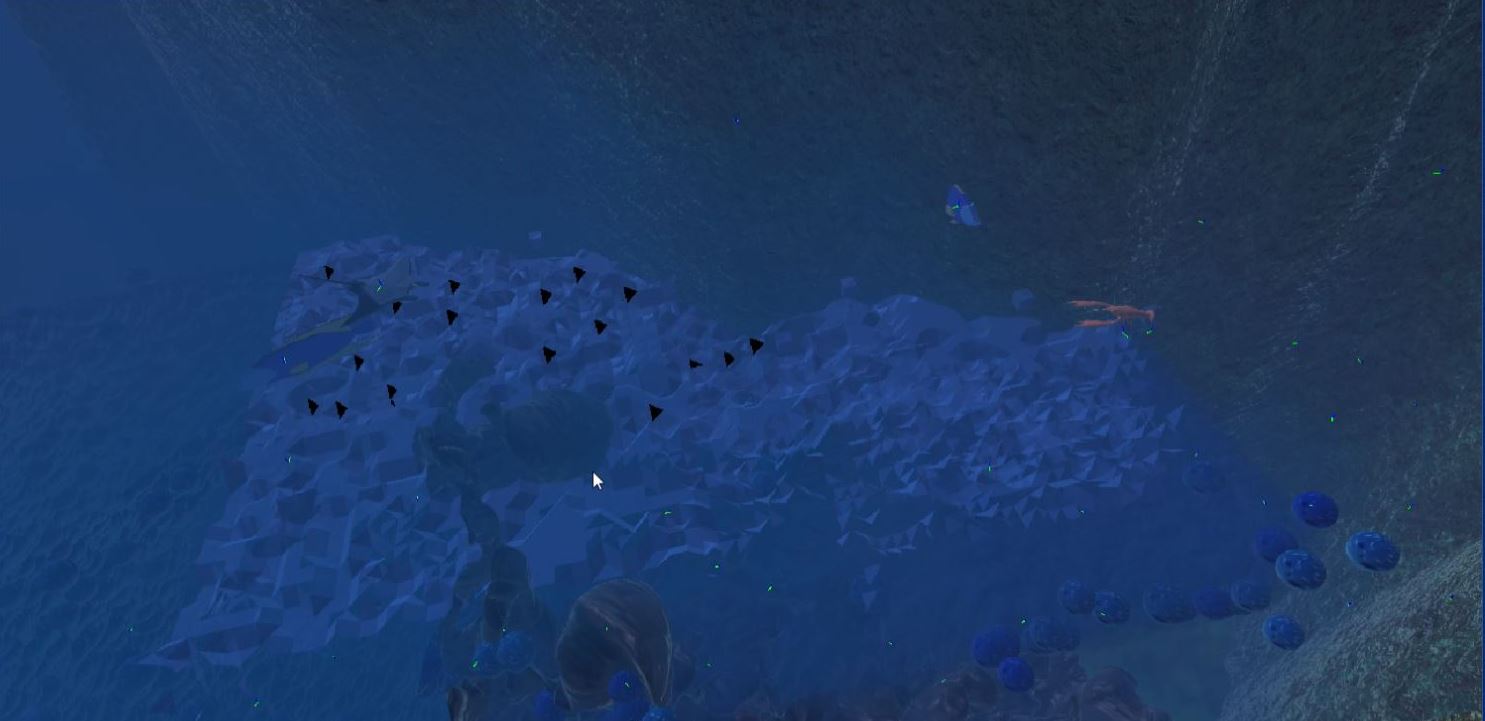}
    \includegraphics[width = 0.23\textwidth, height = 0.15\textwidth]{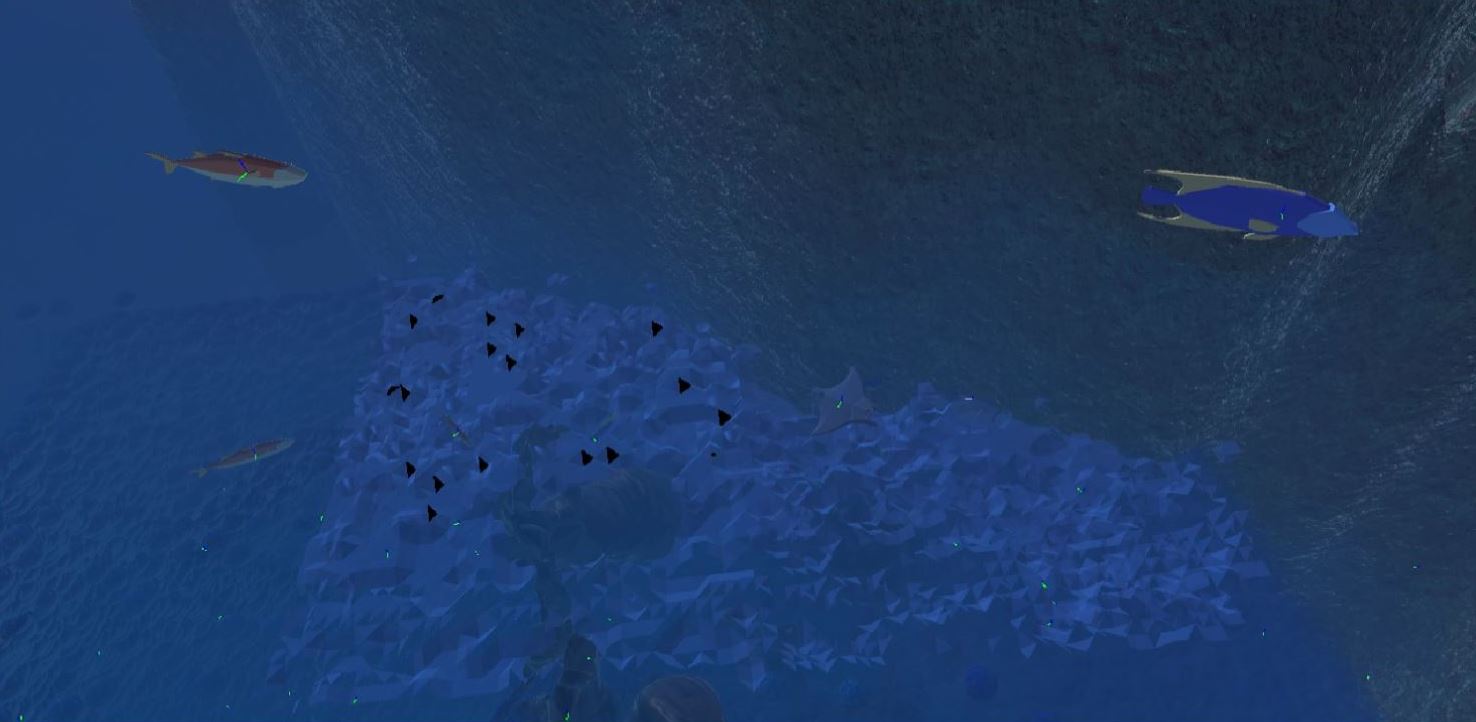}
    \caption{parallel marching cubes}
    \label{parallel marching cubes}
\end{figure}

If every other particle is considered as the neighbor of a particle, then it has a high time space complexity of $O(n ^ 2)$. Therefore, we divide the global problem into smaller parallelizable subproblems, only consider the nearby particles within a small distance as the neighbors of a given particle. The parallel marching cubes algorithm for water mesh construction shares the same parallel neighborhood search algorithm with the SPH and boid algorithms. The parallel marching cubes algorithm extracts the isosurface from voxel density computed from particle density. See figure \ref{parallel marching cubes}.

For each particle in a given time, we compute the cell that the particle belongs to. We set the boundaries of the particles with a set $bound$ = \{min$_x$, max$_x$, min$_y$, max$_y$, min$_z$, max$_z$\}. Let $(c_x, c_y, c_z) \in \mathbb{Z} ^ {\geq 0}$ be the integer coordinates of the cell that encloses the particle, $p \in \mathbb{R} ^ 3$ be the particle's position, $m = (\min_x, \min_y, \min_z) \in \mathbb{R} ^ 3$ be the minimum positional value of the bounding box, $ma = (\max_x, \max_y, \max_z) \in \mathbb{R} ^ 3$ be the maximum positional value of the bounding box, $d \in \mathbb{R} ^ {\geq 0}$ be the cell size, we define a function $get\_cell: \mathbb{R} ^ 3 \rightarrow \mathbb{Z} ^ 3$
\begin{align}
get\_cell(p) &= \left\lfloor\frac{p - m}{d}\right\rfloor = (c_x, c_y, c_z)
\end{align}

We maintain the number of cells for each dimension in $\dim \in \mathbb{Z}_{\geq 0} ^ 3, \dim = (ma - m) / d$. Let \textit{max\_particles\_per\_grid} $\in$ $\mathbb{Z} ^ {\geq 0}$ be the maximum number of particles that can be in a given cell at a given time. After obtaining the cell number $(c_x, c_y, c_z)$ in which the particle belongs to, we compute a hash from this cell number, the result $hash\_grid\_position \in \mathbb{Z}$ is a hash that uniquely identifies the particle's cell, then we use $hash\_grid\_position \times max\_particles\_per\_grid + original\_value$ as the index into an array $hash\_grid$ of size $(dim_x + dim_x (dim_y + dim_y dim_z)) max\_particles\_per\_grid$ that stores all the particles that belong to each grid, in which the \textit{original\_value} $\in \mathbb{Z}$ is the current count of neighbors that are recorded for this particular cell. We use a spatial hashing function $hash: \mathbb{Z} ^ 3 \rightarrow \mathbb{Z}$ defined by $hash(c_x, c_y, c_z) = c_x + \dim_x(c_y + \dim_y c_z)$.
\begin{lstlisting}[caption = Algorithm compute hash grid]
[numthreads(thread_x, thread_y, thread_z)]
void compute_hash_grid(uint3 id : SV_DispatchThreadID)
    int original_value = 0,
    hash_grid_position = hash(get_cell(particles[id.x].position));
    InterlockedAdd(hash_grid_tracker[hash_grid_position], 1, original_value);
    hash_grid[hash_grid_position * max_particles_per_grid + original_value] = id.x;
\end{lstlisting}

After storing the particle indices (index to query from the $particles$ array) that each cell have into $hash\_grid$ in \textit{compute\_hash\_grid} function, we compute the neighbor list for each particle. Let $position \in \mathbb{R} ^ 3$ be the position of a given particle, $origin\_index = \lfloor (position - m) / d \rfloor \in \mathbb{Z} ^ 3$. We define a function $get\_nearby\_key$ to compute the $nearby\_key$ array that stores the hashed one dimensional indices of cells ($hash(c_x, c_y, c_z)$) that are near the particle at $position$. We loop through the $3 ^ 3 = 27$ cells that are near the particle, the center of these 27 cells is the cell that contains this particle.
\begin{lstlisting}[caption = Algorithm get nearby key]
void get_nearby_key(int3 origin_index, float3 position, out int nearby_key[27])
    int3 nearby_index[27];
    int idx = 0;
    for(int i = -1; i <= 1; ++i)
        for(int j = -1; j <= 1; ++j)
            for(int k = -1; k <= 1; ++k)
                nearby_index[idx++] = origin_index + int3(i, j, k);

    for(uint a = 0; a < 27; ++a)
        int3 cell = nearby_index[a];
        if(cell.x < 0 || cell.x >= dimension.x || cell.y < 0 || cell.y >= dimension.y || cell.z < 0 || cell.z >= dimension.z)
            nearby_key[a] = -1;
        else
            nearby_key[a] = hash(nearby_index[a]);
\end{lstlisting}

We define a function $compute\_neighbor\_list$ to compute the neighbor indices (index into $particles$ array) for each particle. Let $g = \lfloor (position - m) / d \rfloor \in \mathbb{Z} ^ 3$ be the cell number vector of the cell that contains this particle. We allocate a 27-element integer array $grids$ on the stack to store the hashed one dimensional indices of cells computed from $get\_nearby\_key$. In the outer loop $\forall i \in [0, 27) \cap \mathbb{Z}$, we iterate through the 27 possible cells that are considered nearby cells for this particle. In the inner loop $\forall j \in [0, hash\_grid\_tracker[grid[i]]) \cap \mathbb{Z}$, we traverse through every particle in grid $i$, since grid $i$ was computed as a nearby grid relative to the origin particle, every particle in grid $i$ is considered a $potential\_neighbor$ of the origin particle. The index of $potential\_neighbor$ is computed as $hash\_grid[grids[i] * max\_particles\_per\_grid + j]$ in which it generates the permutation of \textit{hash\_grid\_tracker[grid[i]]} particle indices for a fixed grid number $grid[i]$. Then we calculate the squared Euclidean distance between the potential neighbor with the origin particle. If the distance is less than the smoothing length squared, then we found a neighbor for the origin particle and store the index of the newly found neighbor into $neighbor\_list$.
\begin{lstlisting}[caption = Algorithm compute neighbor list]
[numthreads(thread_group_size, 1, 1)]
void compute_neighbor(uint3 id : SV_DispatchThreadID)
    if(id.x >= n_particle) return;
    particles[id.x].neighbor = 0;
    int3 g = get_cell(particles[id.x].position);
    int nearby_cells[27];
    get_nearby_key(g, particles[id.x].position, nearby_cells);
    for(uint i = 0; i < 27; ++i)
        if(nearby_cells[i] == -1) continue;
        for(uint j = 0; j < min(hash_grid_tracker[nearby_cells[i]], max_particles_per_grid); ++j)
            const uint potential_neighbor = hash_grid[nearby_cells[i] * max_particles_per_grid + j];
            if(potential_neighbor == id.x) continue;
            const float3 v = particles[potential_neighbor].position - particles[id.x].position;
            if(dot(v, v) < radius2)
                neighbors[id.x * max_particles_per_grid * 27 + particles[id.x].neighbor++] = potential_neighbor;
\end{lstlisting}
\section{Neighbor Search with Sorting}
The neighbor search with sorting is computed by first computing the partial sum of particle counts in each cell and use this as the starting position of each cell to find the potential neighbors, then record the count of neighbors for each particle, then compute a partial sum on the neighbor counts to find the neighbor offsets for each particle (index into neighbor array to record and query the neighbors for each particle). The partial sum is computed in parallel using the up sweep and down sweep method \cite{ihmsen_2011} \cite{harris_2007}.

The prefix sums operation is a function $f: \mathbb{R} ^ n \rightarrow \mathbb{R} ^ n$ defined by $(a_0, a_1, ..., a_{n - 1}) \mapsto (a_0, (a_0 \oplus a_1), ..., (a_0 \oplus a_1 \oplus ... \oplus a_{n - 1}))$ in which the $\oplus$ operator is an operation such as addition. The reduce operation is a function $g: \mathbb{R} ^ n \rightarrow \mathbb{R} ^ n$ is defined by $(a_0, a_1, ..., a_{n - 1}) \mapsto (a_0 \oplus a_1 \oplus ... \oplus a_{n - 1})$ \cite{blelloch_1993}.

For each particle, we compute a hash value based on its cell id using spatial hashing. After assigning each particle to a cell, we use counting sort to sort the particles by their hash values. Let $cell\_index$ be a field of $particle$ struct that stores the hashed cell index of this particle, $cell\_particle\_counts$ be an array that stores the number of particles in each cell, $particle\_indices\_in\_cells$ be the internal indices of each particle relative to the start of each cell (first particle in this cell have 0, second particle in this cell have 1, ...), $cell\_offsets$ be an array containing the starting indices of each cell in the sorted array, which is computed as the cumulative sum of the number of particles in the cells inside and before each cell queried, $sort\_indices$ be an array that stores the destination sorted indices sorted by each particle's one dimensional cell index.
\begin{lstlisting}[caption = Algorithm count sort particle index]
void set_particle_count(particle * particles, uint * cell_particle_counts, uint * particle_indices_in_cells)
    uint id = blockIdx.x * blockDim.x + threadIdx.x,
    cell_index = hash(get_cell(particles[id]));
    particles[id].cell_index = cell_index;
    particle_indices_in_cells[id] = atomicAdd(&cell_particle_counts[cell_index], 1);

void count_sort_particle_index(uint * cell_offsets, uint * particle_indices_in_cells, uint * sort_indices)
    uint id = blockIdx.x * blockDim.x + threadIdx.x,
    cell_index = particles[id].cell_index,
    sort_index = cell_offsets[cell_index] + particle_indices_in_cells[id];
    sort_indices[sort_index] = id;

#pragma kernel compute_neighbor_count
[numthreads(thread_per_group, 1, 1)]
void compute_neighbor_count(uint3 id : SV_DispatchThreadID)
    if(id.x >= n_particle) return;
    int3 origin_index = get_cell(particles[id.x].position);
    int count = 0;
    for(int x = -1; x < 2; ++x)
        for(int y = -1; y < 2; ++y)
            for(int z = -1; z < 2; ++z)
                int3 cell = origin_index + int3(x, y, z);
                int cell_index = hash(cell),
                cell_count = cell_particle_counts[cell_index],
                cell_start = cell_offsets[cell_index];
                for(int i = cell_start; i < cell_start + cell_count; ++i)
                    float3 diff = particles[sort_indices[i]].position - particles[id.x].position;
                    float square_distance = dot(diff, diff);
                    if(square_distance < radius2)
                        ++count;
    particles[id.x].neighbor = count;
    neighbor_count[id.x] = count;
\end{lstlisting}

We count the number of particles for each cell using an atomic increment, we perform a parallel prefix sum (scan) to compute each particle's destination address through $cell\_offsets[cell\_index] + particle\_indices\_in\_cells[id]$ \cite{Hoetzlein_2014} \cite{Green_2010} \cite{Groß_2019}. We then write the particles to contiguous locations in $sort\_indices$ array.

After obtaining the sorted particle indices inside the $sort\_indices$ array, we compute the count of neighbors for each particle inside the set of $3 ^ 3 = 27$ neighboring cells of the given particle, after obtaining the neighbor counts inside $neighbor\_count$, we perform another parallel prefix sum (scan) to find the partial sum of neighbor counts and store this partial sum into an array $neigbor\_offsets$ \cite{harris_2007} \cite{Band_2019} \cite{Franklin_2006} \cite{Hillis_1986}. Then we find the indices of all neighbors of all particles and store that into $neighbors$ array. Let $count = 0, write\_offset = neighbor\_offset[id.x]$, in which $count$ is the current count of neighbors of the given particle whose indices are already stored into the $neighbors$ array. We use an assignment $neighbors[write\_offset + count] = sort\_particle\_index[i]$ to store the sorted particle index into the $neighbors$ array. The $neighbors$ array is then used in the computation of the density and forces to query the neighbors of each particle. Our approach avoids recomputation of particle neighbors during computation of the density and acceleration step at the expense of space for storing the neighbor offsets and neighbor indices.
\section{Voxel Density}
To perform mesh reconstruction from particles, we need to construct a surface that wraps around the particles given particle positions. The metaball approach is best fit for only a few particles \cite{zhu_2005}. We instead uses the particle neighborhood information for each cube corner. Let $x \in \mathbb{R} ^ 3$ be a cube corner inside the bounds of the SPH simulation, $\overline{x} \in \mathbb{R} ^ 3$ be the weighted average particle position, $\overline{r} \in \mathbb{R}$ be the weighted average particle radius. The voxel density function $\phi: \mathbb{R} ^ 3 \rightarrow \mathbb{R}$ that we used is
\begin{align}
\phi(x) &= |x - \overline{x}| - \overline{r}\\
\overline{x} &= \sum_i w_i x_i, \quad \overline{r} = \sum_i w_i r_i\\
w_i &= \frac{k(|x - x_i| / R)}{\sum_j k(|x - x_j| / R)}
\end{align}
$k: \mathbb{R} \rightarrow \mathbb{R}, k(s) = (1 - s ^ 2) ^ 3$ is the kernel function that we used. $R \in \mathbb{R}$ is the radius of neighborhood around $x$.
\begin{lstlisting}[caption = Algorithm compute voxel density]
[numthreads($thread_x, thread_y, thread_z$)]
void compute_density(uint3 id: SV_DispatchThreadID)
    cube_corner_pos = $(x_{min}, y_{min}, z_{min})$ + id * grid_size $\in \mathbb{R} ^ 3$
    cube_corner_index = id.x + n_point_per_axis * (id.y + n_point_per_axis * id.z) $\in \mathbb{Z}$
    sum = 0, weight_divisor = 0, weight_multiplier $\in \mathbb{R}$
    average_particle_pos = (0, 0, 0) $\in \mathbb{R} ^ 3$
    for i $\in$ [0, cube_corner_neighbor_tracker[cube_corner_index]] $\cap \mathbb{Z}$
        neighbor_index = cube_corner_neighbor_list[cube_corner_index * max_particles_per_cube + i]
        weight_divisor += kernel(length(cube_corner_pos - particles[neighbor_index].position) / radius)
    weight_multiplier = 1 / weight_divisor
    for i $\in$ [0, cube_corner_neighbor_tracker[cube_corner_index]] $\cap \mathbb{Z}$
        neighbor_index = cube_corner_neighbor_list[cube_corner_index * max_particles_per_cube + i]
        x = length(cube_corner_pos - particles[neighbor_index].position) / radius
        weight_i = weight_multiplier * kernel(x)
        average_particle_pos += weight_i * particles[neighbor_index].position
    points[cube_corner_index] = cube_corner_pos
    voxel_density[cube_corner_index] = length(cube_corner_pos - average_particle_pos) - particle_size
\end{lstlisting}
\begin{figure}
    \centering
    \includegraphics[width = 0.49\textwidth, height = 0.3\textwidth]{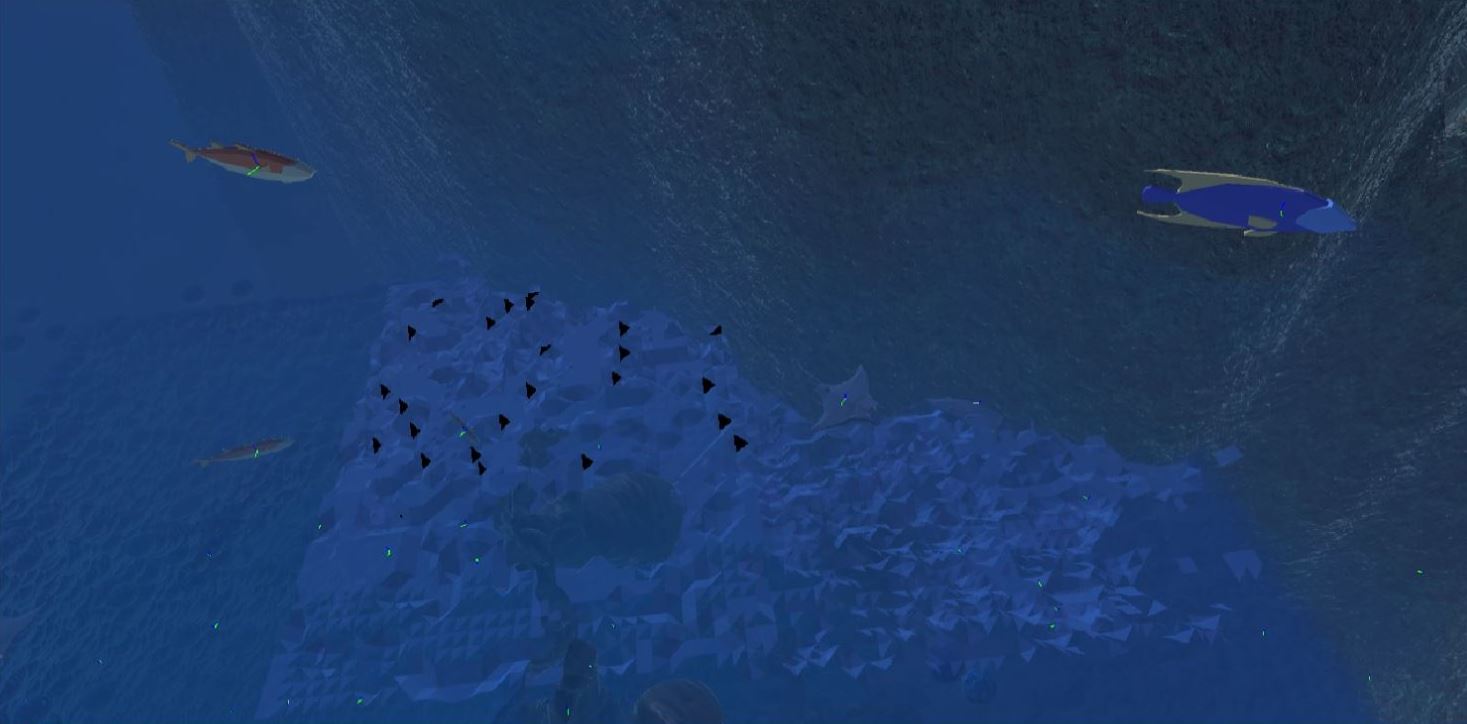}
    \caption{Surface reconstruction from particle attributes underwater using marching cubes. The dark mesh near the middle of the fluid substratum is an underwater cave.}
    \label{surface reconstruction}
\end{figure}
\section{Marching Cubes}
The marching cubes algorithm locates the surface, creates triangles, and compute the normals of the surface for each vertex of the triangles. For each cube, the algorithm computes the intersection representation between the cube and the surface. The surface intersections are computed by comparing the voxel value at the vertex with the isovalue of the surface. If the voxel value at the vertex is below the isovalue, we assign a one to the cube's vertex. If the voxel value at the vertex is greater than or equal to the isovalue, we assign a zero to the cube's vertex.
\begin{lstlisting}[caption = Algorithm cube index]
cube_index = 0;
for(int i = 0; i < 8; ++i)
    if(cube_corner_density[i] < isolevel)
        cube_index |= 1 << i;
\end{lstlisting}
\begin{figure}
    \centering
    \includegraphics[width = 0.15\textwidth, height = 0.15\textwidth]{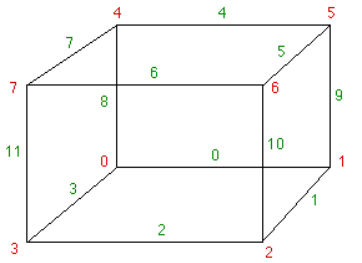}
    \includegraphics[width = 0.15\textwidth, height = 0.15\textwidth]{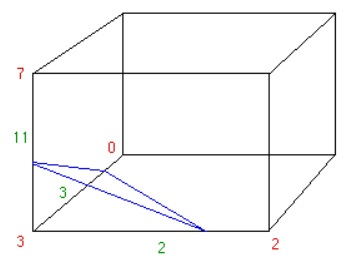}
    \caption{marching cubes}
    \label{marching cubes}
\end{figure}

Consider the configuration in figure \ref{marching cubes} \cite{bourke_1994}, in which green labels denote edge indices and red labels denote vertex indices, suppose the voxel value at vertex 3 is less than the isovalue and every other vertices have a voxel value greater than the isovalue. By algorithm cube index, we obtain a $cube\_index = (0 | (1 << 3))_2 = (1000)_2 = (8)_{10}$. Then, we use the $cube\_index$ as the array index of $edge\_table$, finding that $edge\_table[8] = (80c)_{16} = (1000\,0000\,1100)_2$, since bits at position 2, 3, 11 are ones, we find that the surface intersects with cube at the edges 2, 3, and 11.

We use linear interpolation to compute the vertex positions that are at the intersecting edges. Let $i \in \mathbb{R}$ be the isovalue, $p_1, p_2 \in \mathbb{R} ^ 3$ be vertices of a intersecting edge, $v_1, v_2$ be voxel values at each vertex, then the vertex position $p \in \mathbb{R} ^ 3$ at the intersecting edge is
\begin{align}
p &= p_1 + \frac{i - v_1}{v_2 - v_1}(p_2 - p_1)
\end{align}

To find the triangle facets from vertex positions of intersecting edges, we use the same $cube\_index$ as the array index of $triangle\_table$, we find $triangle\_table[8]$ = 
\{3, 11, 2, -1, -1, -1, -1, -1, -1, -1, -1, -1, -1, -1, -1, -1\}. By algorithm triangle construction, we construct the triangle from vertices 3, 11, and 2.
\begin{lstlisting}[caption = Algorithm triangle construction]
for(int i = 0; triangulation[cube_index][i] != -1; i += 3)
    triangle a;
    a.vertex_a = vertex_list[triangle_table[cube_index][i]];
    a.vertex_b = vertex_list[triangle_table[cube_index][i + 1]];
    a.vertex_c = vertex_list[triangle_table[cube_index][i + 2]];
    triangles.append(a);
\end{lstlisting}
\begin{figure}
    \centering
    \includegraphics[width = 0.15\textwidth, height = 0.1\textwidth]{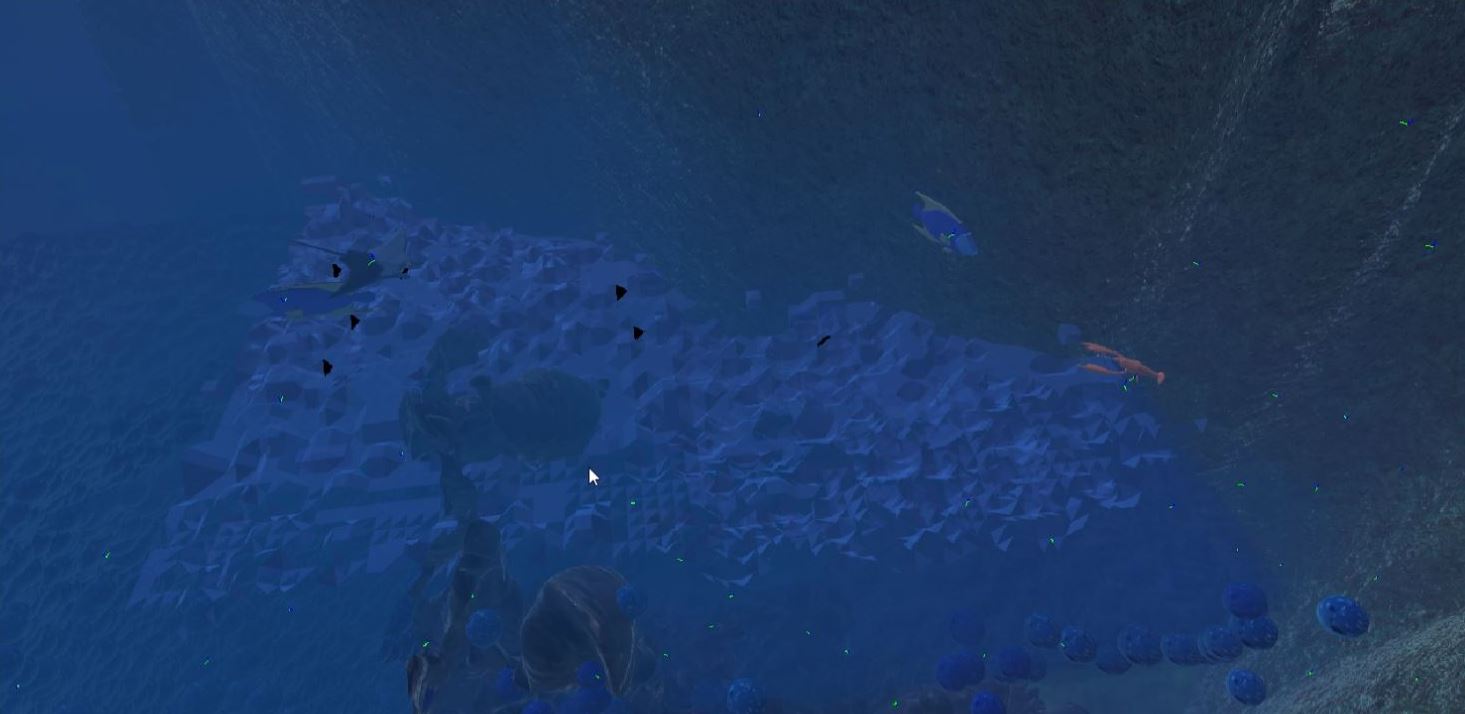}
    \includegraphics[width = 0.15\textwidth, height = 0.1\textwidth]{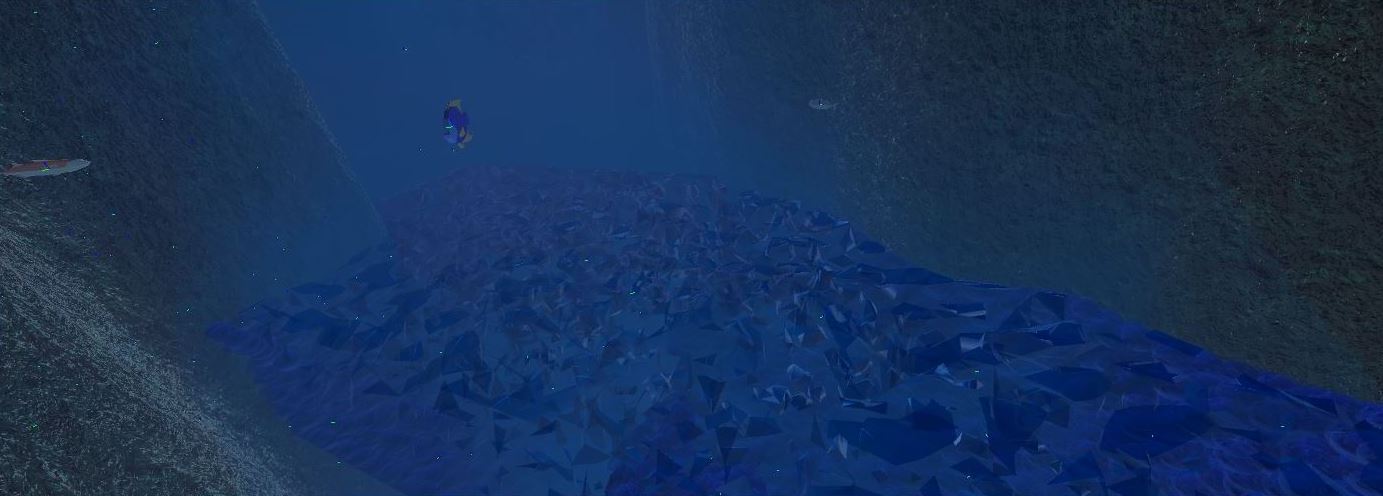}
    \includegraphics[width = 0.15\textwidth, height = 0.1\textwidth]{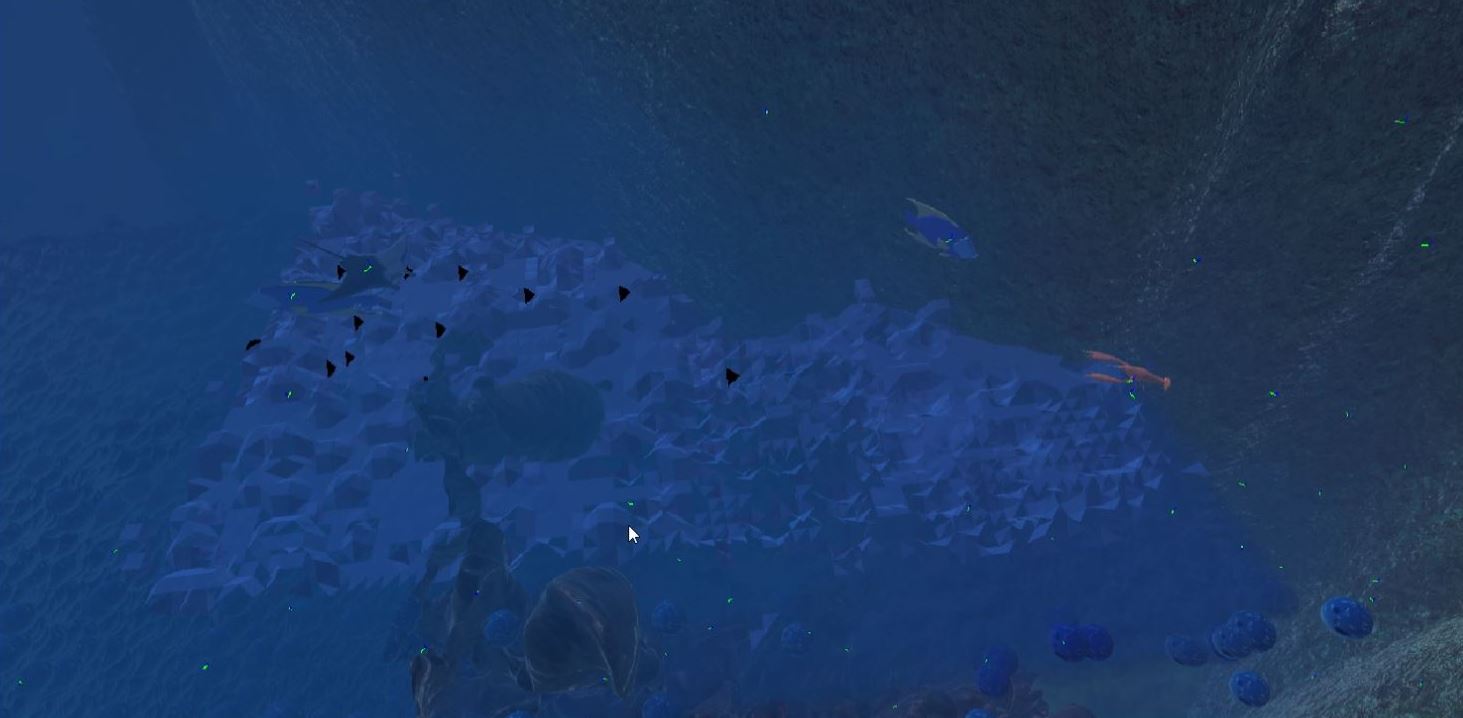}
    \includegraphics[width = 0.15\textwidth, height = 0.1\textwidth]{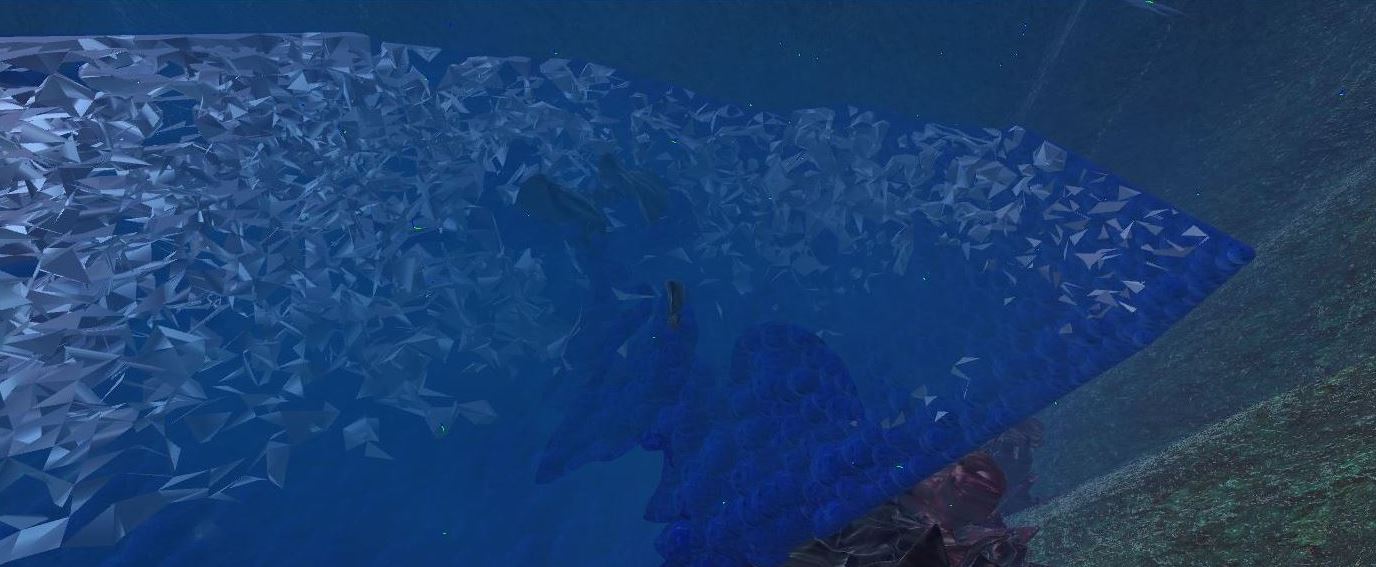}
    \includegraphics[width = 0.15\textwidth, height = 0.1\textwidth]{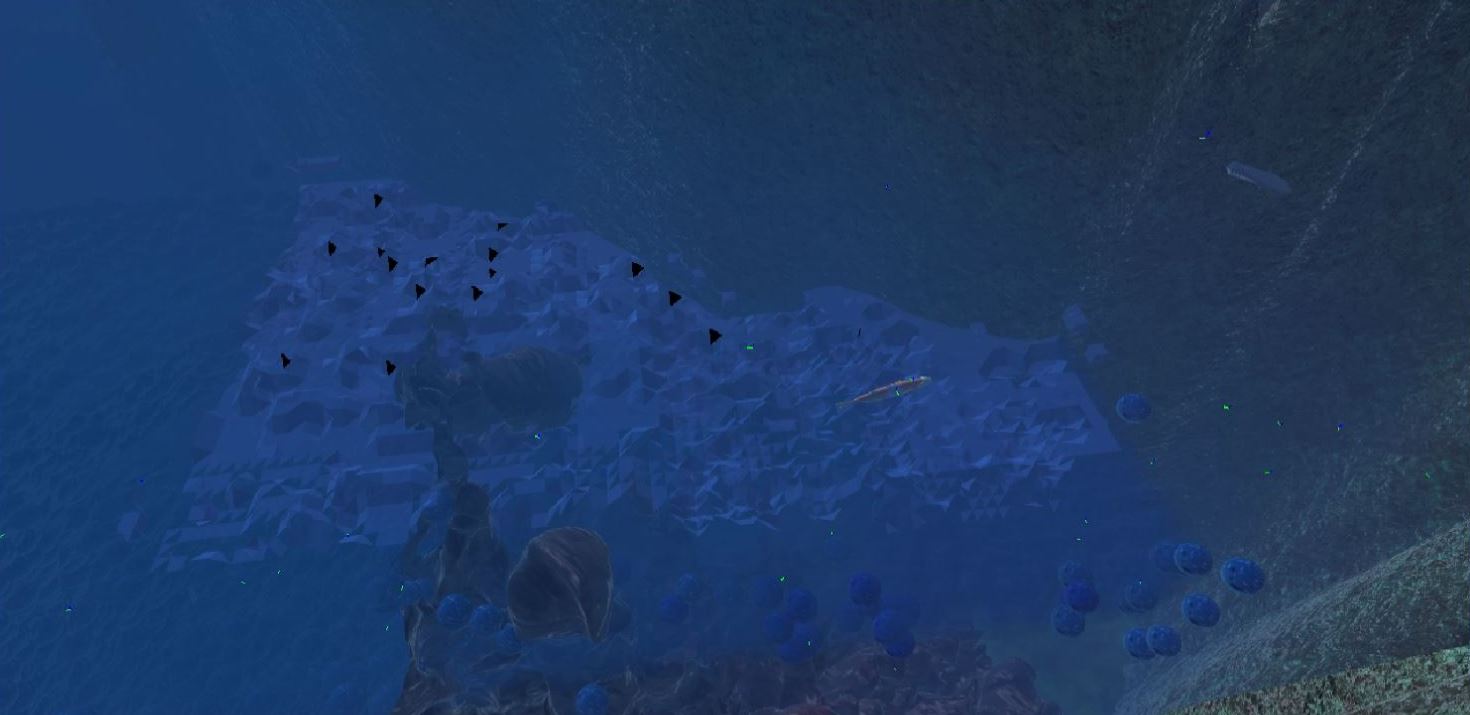}
    \includegraphics[width = 0.15\textwidth, height = 0.1\textwidth]{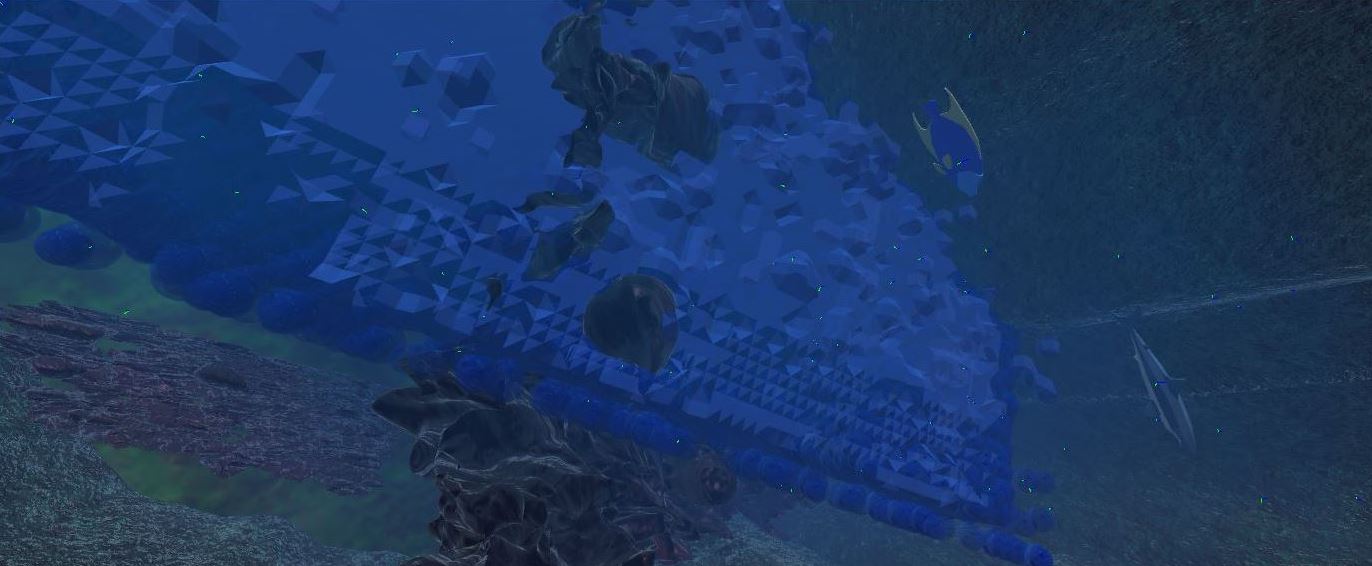}
    \includegraphics[width = 0.15\textwidth, height = 0.1\textwidth]{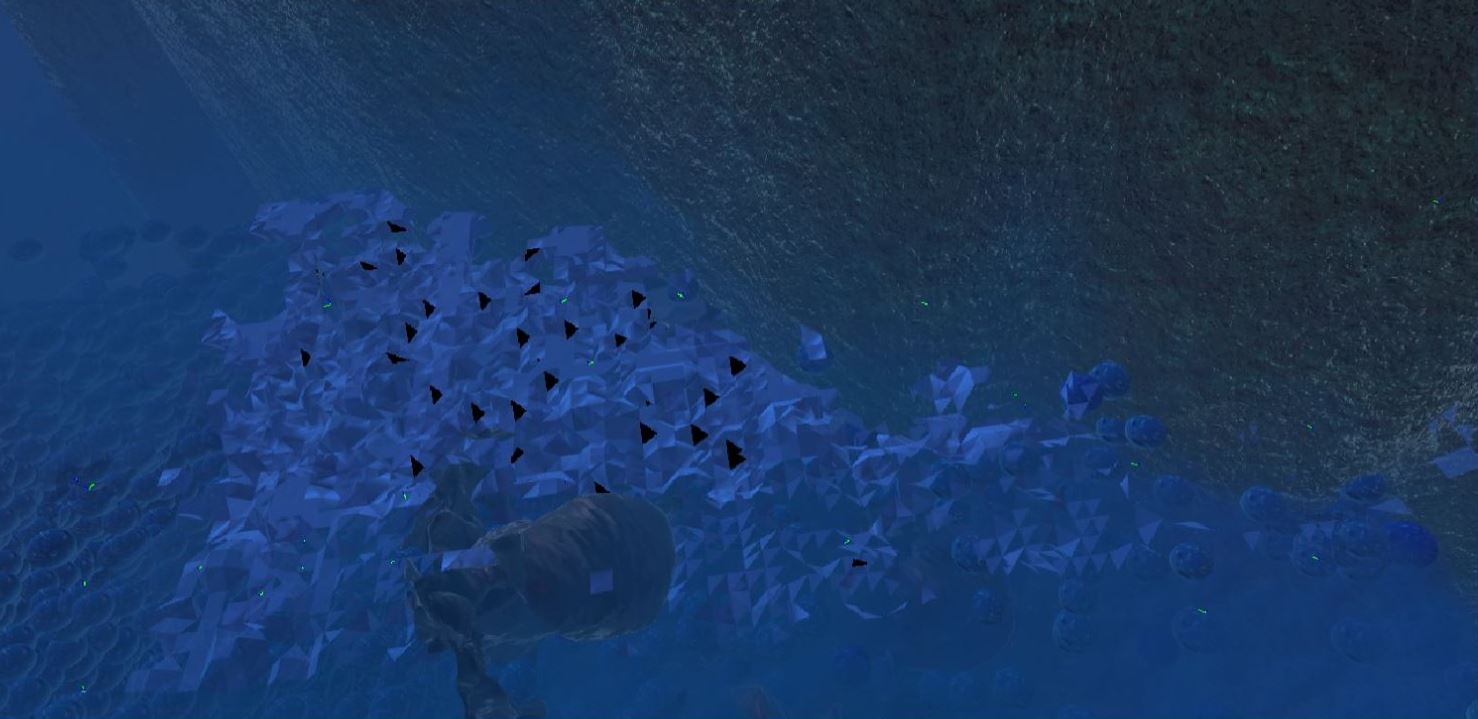}
    \includegraphics[width = 0.15\textwidth, height = 0.1\textwidth]{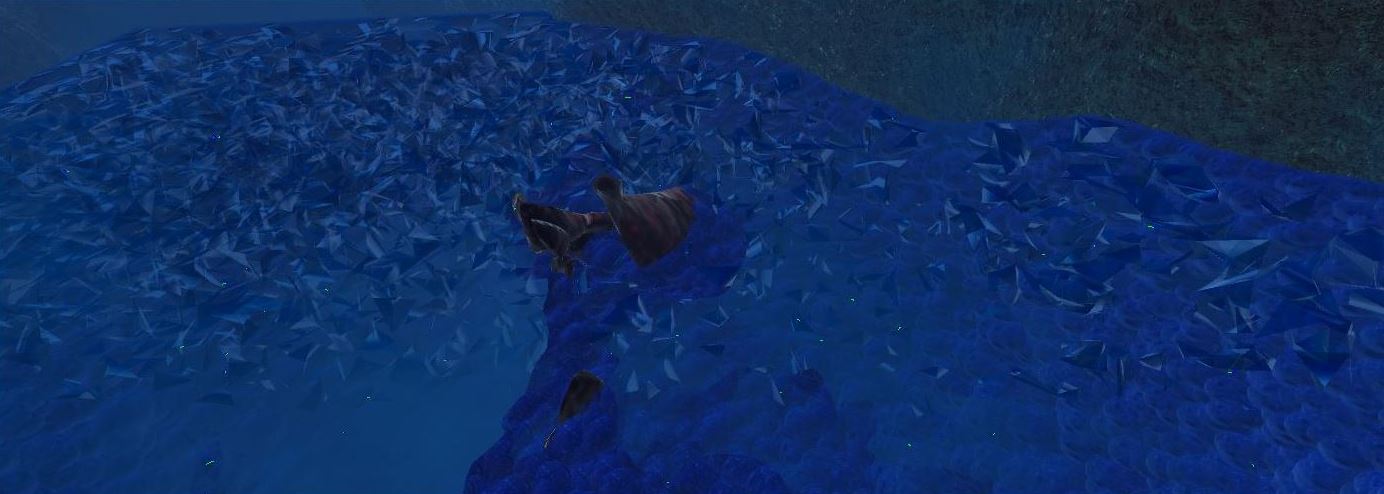}
    \includegraphics[width = 0.15\textwidth, height = 0.1\textwidth]{image/mc_5.jpg}
    \caption{The figures show the different triangle configurations produced by the parallel marching cubes algorithm.}
    \label{different triangle}
\end{figure}
\section{Experimental Results}
The GPU used to perform the experiments is Nvidia GeForce GTX 1080 with Pascal microarchitecture, integrated RAMDAC, and 60Hz refresh rate. The number of CUDA cores used is 2560 with a 1607 MHz graphics clock, 10.01 Gbps memory data rate, 256-bit memory interface, and 320.32 GB/s memory bandwidth. The total available graphics memory is 24400 MB, dedicated video memory is 8192 MB, and shared system memory is 16208 MB. The CPU used is an Intel Xeon W-2123 Processor with 31.7 GB installed usable RAM and a 3.6 GHz processor base frequency. While the experiments performed mainly use the computational capacity of the GPU, the parallel neighbor search algorithms used involve complex calculations including the up sweep and down sweep method that can be efficiently executed on multicore CPUs, which are well-suited for branching, conditional statements, and irregular data access patterns. CPUs can handle these aspects better than GPUs, which excel at massively parallel, regular computations.

\begin{figure}[h!]
    \centering
    \includegraphics[width = 0.23\textwidth, height = 0.16\textwidth]{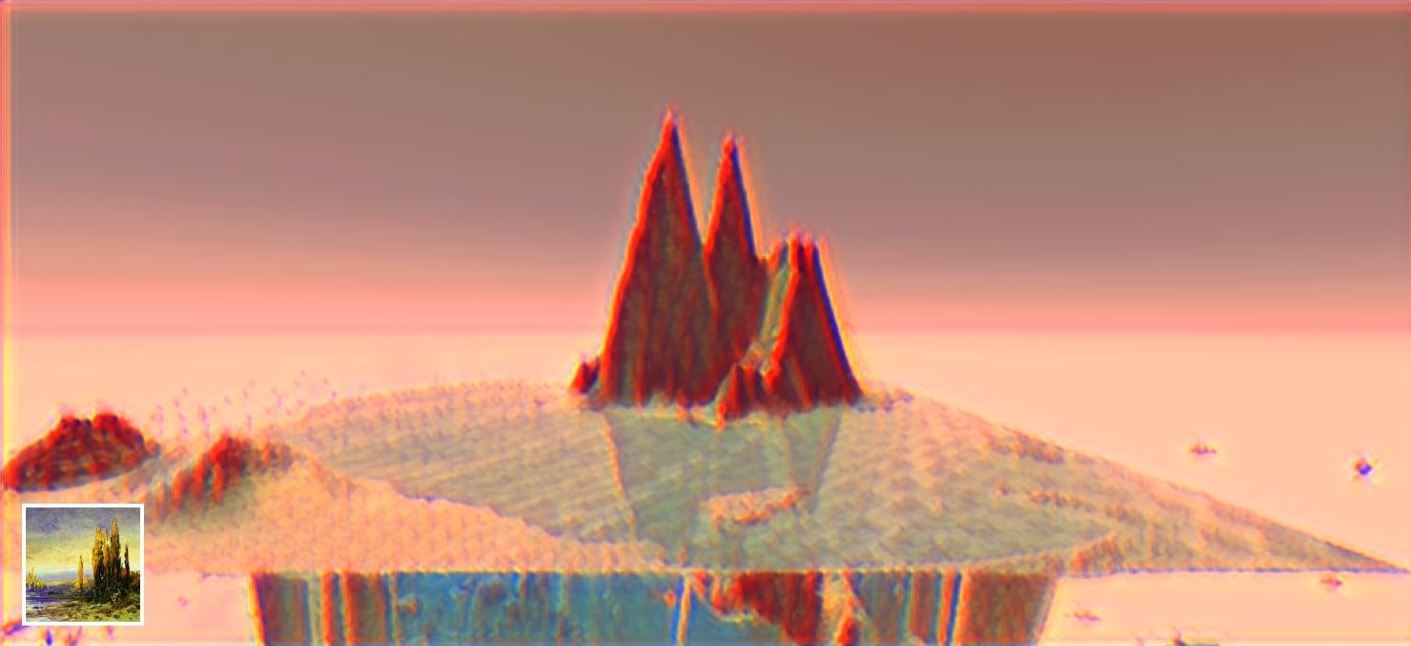}
    \includegraphics[width = 0.23\textwidth, height = 0.16\textwidth]{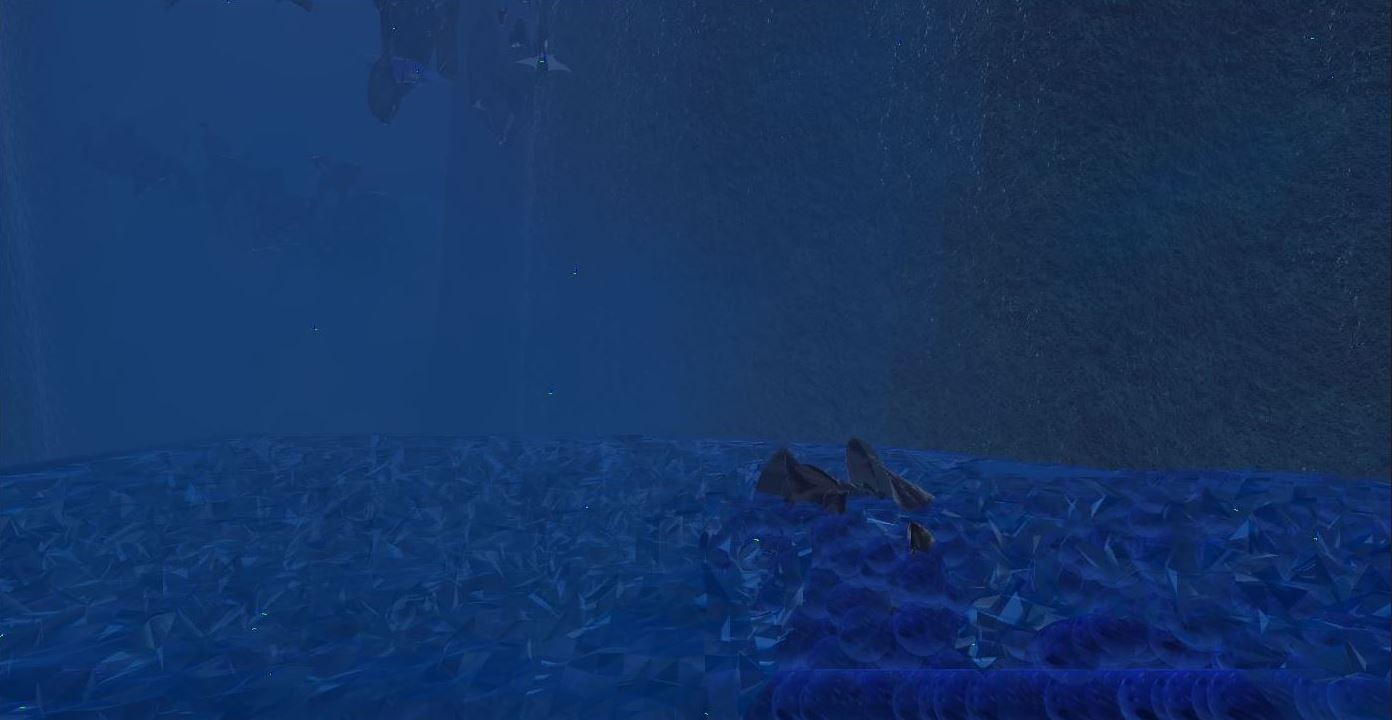}
    \includegraphics[width = 0.23\textwidth, height = 0.16\textwidth]{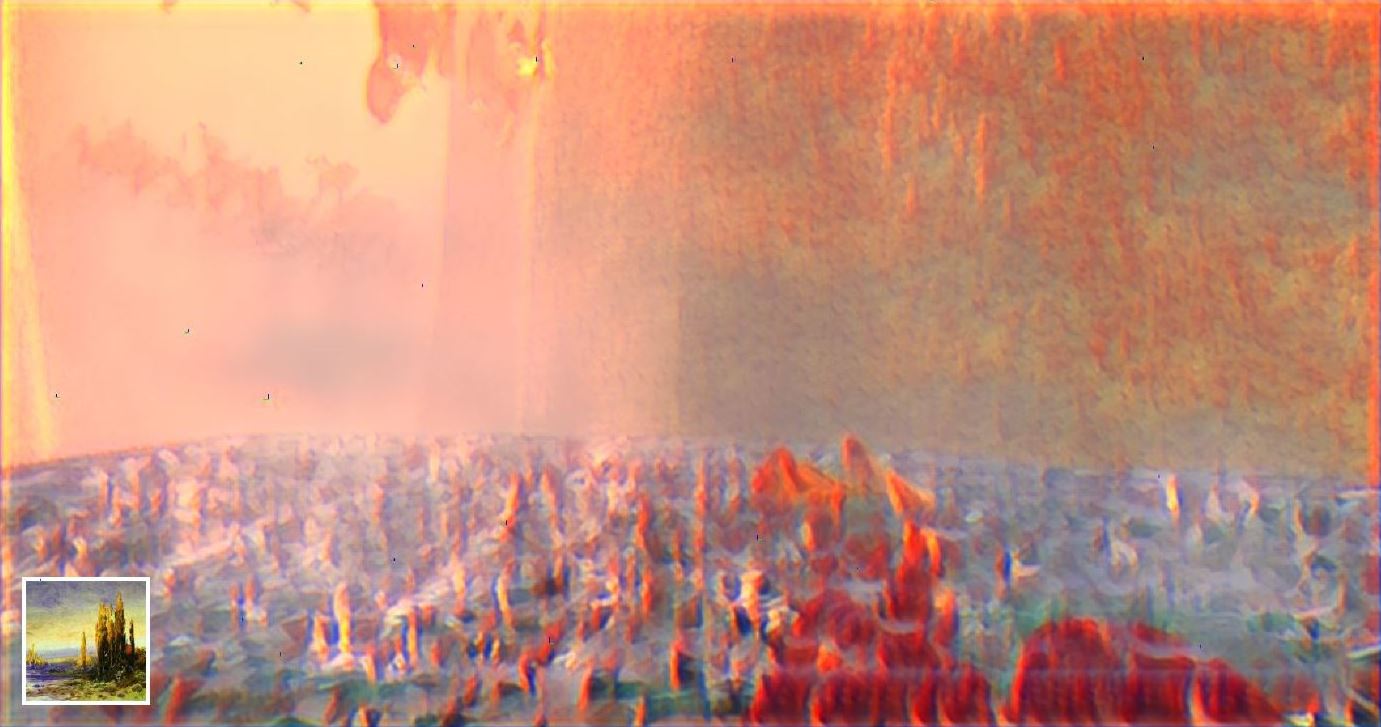}
    \includegraphics[width = 0.23\textwidth, height = 0.16\textwidth]{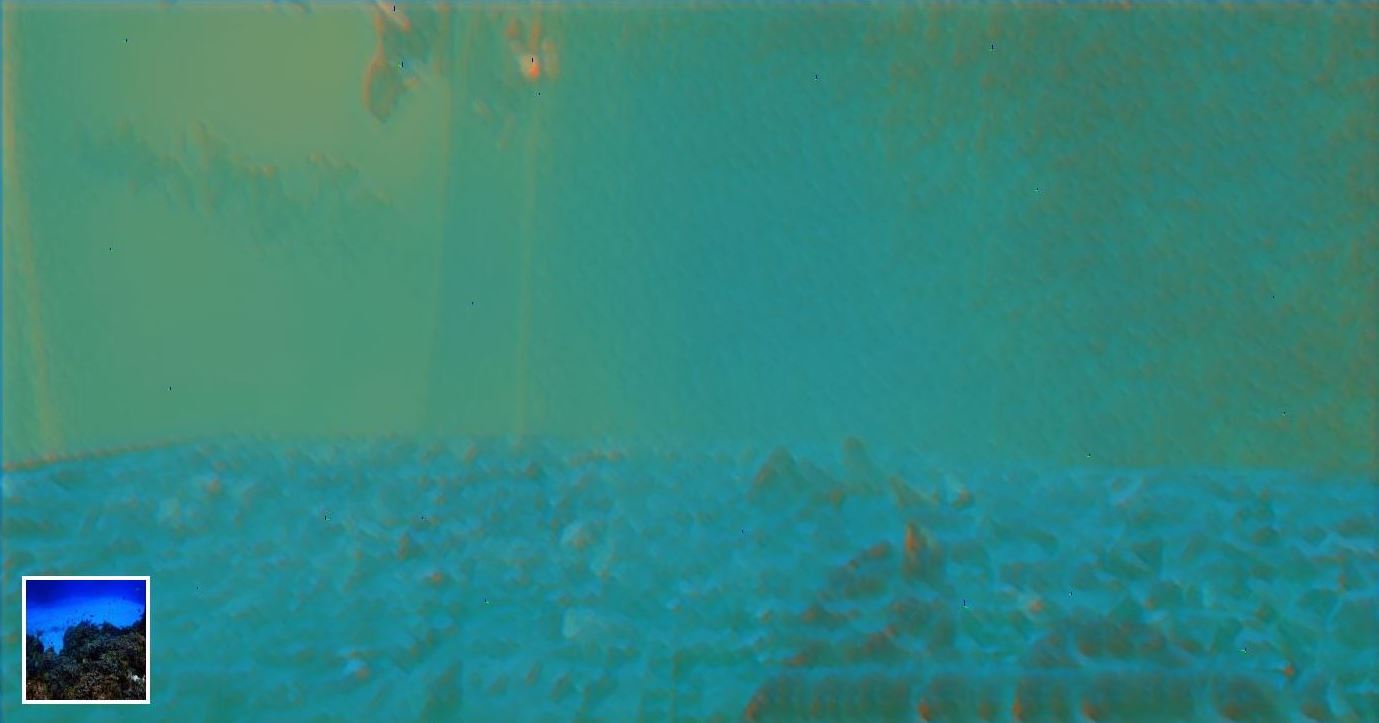}
    \caption{neural post processing}
    \label{style}
\end{figure}

We performed a set of experiments measuring the change of neighbor search time and average number of neighbors per particle in SPH and marching cubes with respect to changes in particle count, grid size, smoothing radius length, and threads per group, in which the results are reported in table \ref{SPH neighbor search time}, \ref{marching cubes neighbor search time}. We observed that the neighbor search time, marching cubes time, and average number of neighbors per particle increase nonlinearly with respect to increase in particle count. The particle radius is chosen to be 0.5 in the simulation units. The length of the cubic grid is chosen to be approximately four times the particle radius. If the grid size is too big, then a given particle can find too many neighbors, making its iterative forward update very slow. If the grid size is too small, then there are not enough neighbors found for a given particle to enable observable hydrodynamical effects. We also applied neural style transfer per frame using Unity’s Barracuda library in the post-processing stage to the voxel density fluid mesh to complement the surface shading step after the marching cubes stage. See figure \ref{style}. The algorithm provides a reasonable fluid-agent interaction prediction in mesoscale, which can deviate from real-world behavior due to unaccounted interactions at the microscopic level, in which molecular dynamical models can be used. For a finer grain simulation, the microfluidic dynamics should involve the modeling of individual molecules of the aqueous solution taking into account intramolecular, intermolecular, supramolecular, van der Waals, electromagnetic, nuclear, and thermal effects and forces.

The assignment of multiple particles to each GPU thread is often done to improve the computational efficiency of simulations. By grouping multiple particles per thread, we can reduce the overhead of thread management and synchronization. However, the optimal number of particles per thread may vary depending on the specific simulation and hardware configuration. Increasing the number of particles per thread could potentially increase the performance of the algorithm by utilizing the GPU's parallel processing capabilities more effectively, but there's a trade-off between the number of particles and the memory available on each thread. An excessively large number of particles per thread could lead to memory limitations and reduced performance. Finding the right balance typically involves experimentation and performance profiling to determine the optimal configuration for a given simulation. The utilization of multiple CPUs and GPUs can significantly accelerate the performance of our developed simulator, especially when dealing with complex computational tasks and intricate fluid dynamics simulations. By employing a distributed computing architecture that incorporates multiple CPUs and GPUs, we can parallelize various aspects of the simulation process, effectively dividing the computational workload across these resources. This parallelization approach not only enhances the overall processing power but also enables us to handle larger datasets and perform more detailed simulations, thereby facilitating a more accurate representation of fluid dynamics, particularly at the molecular level. The utilization of multiple CPUs and GPUs empowers the efficient modeling of intricate molecular dynamics and complex forces governing fluid particles and bioorganic agents, facilitating a comprehensive exploration of the underlying molecular mechanisms at the heart of fluidic environments. This advanced computing architecture not only accelerates simulations but also fosters a deeper understanding of the nuanced interactions at the molecular level, enhancing our insights into the intricate dynamics of viscoelastic media and bioorganic systems.

\begin{figure}[h!]
    \centering
    \includegraphics[width = 0.49\textwidth, height = 0.3\textwidth]{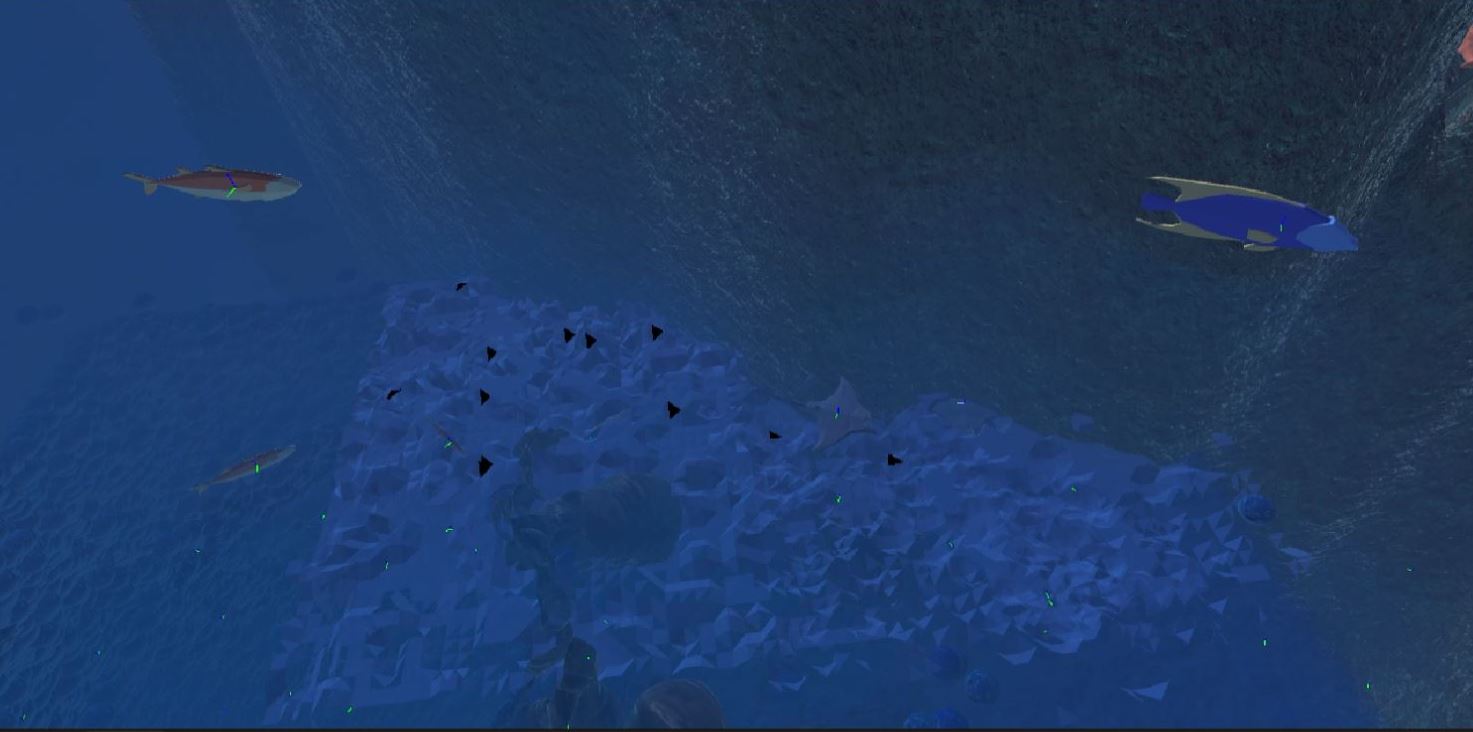}
    \caption{parallel marching cubes voxel density reconstruction experiments}
    \label{voxel density}
\end{figure}

Currently the smoothed particle hydrodynamics simulation, boids, and the marching cubes voxel density reconstruction pipeline are implemented in the compute shader. For high-fidelity simulation and details in the fluid mesh, we need a high amount of particles. When the particle count is over $10 ^ 6$, the simulation becomes too slow to be run. In our implementation, the neighbor search with sorting has a lower space-time complexity than the neighbor search without sorting. For the marching cubes aspect, there are many cases the triangles on the particles are only partially constructed, this might be due to too much computation is assigned for each frame and the GPU is not able to finish the computations. Out-of-memory errors were often encountered. When the number of particles is too small, there are not enough triangles constructed from the marching cubes algorithm for the triangles to connect with each other. See figure \ref{parallel marching cubes}, \ref{surface reconstruction}, \ref{different triangle}, \ref{voxel density}. Distributed multicore CPUs can be used to complement GPUs in our smoothed particle hydrodynamical and interactive boid simulator using the Message Passing Interface (MPI) or Open Multi-Processing (OpenMP) to help mitigate memory limitations and improve simulation speed. When each GPU thread per particle is not assigned with a sufficient amount of work, the cost of communication between threads outweighs the parallelization benefit achieved as compared with assigning the GPU thread with multiple particles or using a distributed multicore CPU hybrid approach. SPH and boid simulations require storing and managing large datasets, including particle positions, densities, pressures, forces, neighbor lists, and other state attributes. CPUs can be better suited for flexible data management tasks, allowing for efficient handling of particle data and reducing the need for constant data transfers between the CPU and GPU memory, which can be a bottleneck. By offloading specific tasks to CPU cores, we can use a hybrid parallelization approach, utilizing both CPU and GPU resources effectively.

\begin{table}
\caption{SPH neighbor search time}
\label{SPH neighbor search time}
{\footnotesize
\begin{tabular}{|c|c|c|c|c|c|}
\hline
\begin{tabular}{@{}c@{}}number\\ of\\particles\end{tabular} & \begin{tabular}{@{}c@{}}grid\\size\end{tabular} & \begin{tabular}{@{}c@{}}smoothing\\radius\\length\end{tabular} & \begin{tabular}{@{}c@{}}threads\\per\\group\end{tabular} & \begin{tabular}{@{}c@{}}average\\number\\of\\neighbors\\per\\particle\end{tabular} & \begin{tabular}{@{}c@{}}neighbor\\search\\time\\ (seconds)\end{tabular}\\\hline
$2.5 \times 10 ^ 4$ & 2.2 & 0.9 & 128 & 52 & $10 ^ {-4}$\\\hline
$5 \times 10 ^ 4$ & 2.1 & 0.8 & 256 & 78 & $2.2 \times 10 ^ {-4}$\\\hline
$10 ^ 5$ & 2 & 0.75 & 512 & 82 & $5 \times 10 ^ {-4}$\\\hline
$2 \times 10 ^ 5$ & 1.9 & 1 & 1024 & 84 & $3.3 \times 10 ^ {-3}$\\\hline
$3 \times 10 ^ 5$ & 1.8 & 2 & 32 & 86 & $3.6 \times 10 ^ {-3}$\\\hline
$7 \times 10 ^ 5$ & 1.7 & 2.1 & 64 & 103 & $4.4 \times 10 ^ {-3}$\\\hline
$10 ^ 6$ & 2.3 & 2.2 & 128 & 114 & $5.1 \times 10 ^ {-3}$\\\hline
\end{tabular}
}
\end{table}
\begin{table}
\caption{marching cubes neighbor search time}
\label{marching cubes neighbor search time}
{\footnotesize
\begin{center}
\begin{tabular}{|c|c|c|c|c|}\hline
\begin{tabular}{@{}c@{}}number of\\particles\end{tabular} & grid size & \begin{tabular}{@{}c@{}}threads\\per\\group\end{tabular} & \begin{tabular}{@{}c@{}}marching\\cubes\\time\\(seconds) \end{tabular} & \begin{tabular}{@{}c@{}}percent of\\connected\\ triangles \end{tabular}\\\hline
$10 ^ 4$ & 2.1 & 64 & $5 \times 10 ^ {-3}$ & 0\%\\\hline
$3 \times 10 ^ {5}$ & 2 & 128 & $3.1 \times 10 ^ {-2}$ & 70\%\\\hline
$4 \times 10 ^ {5}$ & 1.9 & 256 & $3.38 \times 10 ^ {-2}$ & 85\%\\\hline
$5 \times 10 ^ {5}$ & 1.8 & 512 & $4.33 \times 10 ^ {-2}$ & 95\%\\\hline
$7 \times 10 ^ {5}$ & 2.2 & 32 & $10 ^ {-1}$ & 100\%\\\hline
$10 ^ 6$ & 2 & 64 & $1.4 \times 10 ^ {-1}$ & \begin{tabular}{@{}c@{}}100\%\\(less detailed)\end{tabular}\\\hline
\end{tabular}
\end{center}
}
\end{table}
The statistics of the rover agent including its cumulative reward, episode length, policy loss, and value loss are included below in figure \ref{rover}. The cumulative reward, which represents the average cumulative episode reward across all agents, increased, indicating successful training sessions. The episode length is the mean length of each episode in the environment for all agents. The policy loss, indicating the average loss in the policy function update, decreased, indicating less significant changes in the policy. The value loss, reflecting the mean loss in the value function update, decreased, indicating improved predictive accuracy in estimating state values.
\begin{figure}[h!]
    \centering
    \includegraphics[width = 0.23\textwidth, height = 0.16\textwidth]{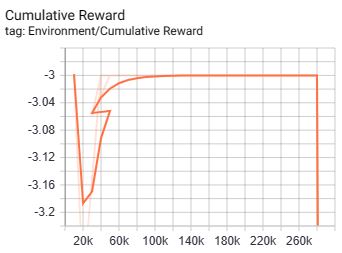}
    \includegraphics[width = 0.23\textwidth, height = 0.16\textwidth]{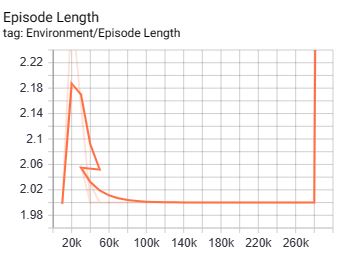}
    \includegraphics[width = 0.23\textwidth, height = 0.16\textwidth]{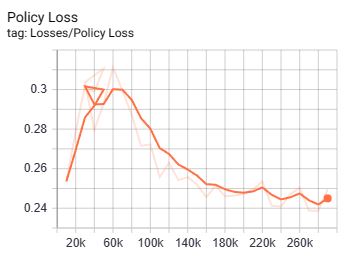}
    \includegraphics[width = 0.23\textwidth, height = 0.16\textwidth]{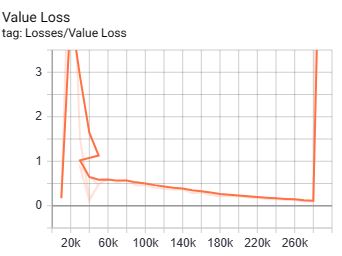}
    \caption{Rover statistics}
    \label{rover}
\end{figure}
\begin{figure}[h!]
    \centering
    \includegraphics[width = 0.49\textwidth, height = 0.3\textwidth]{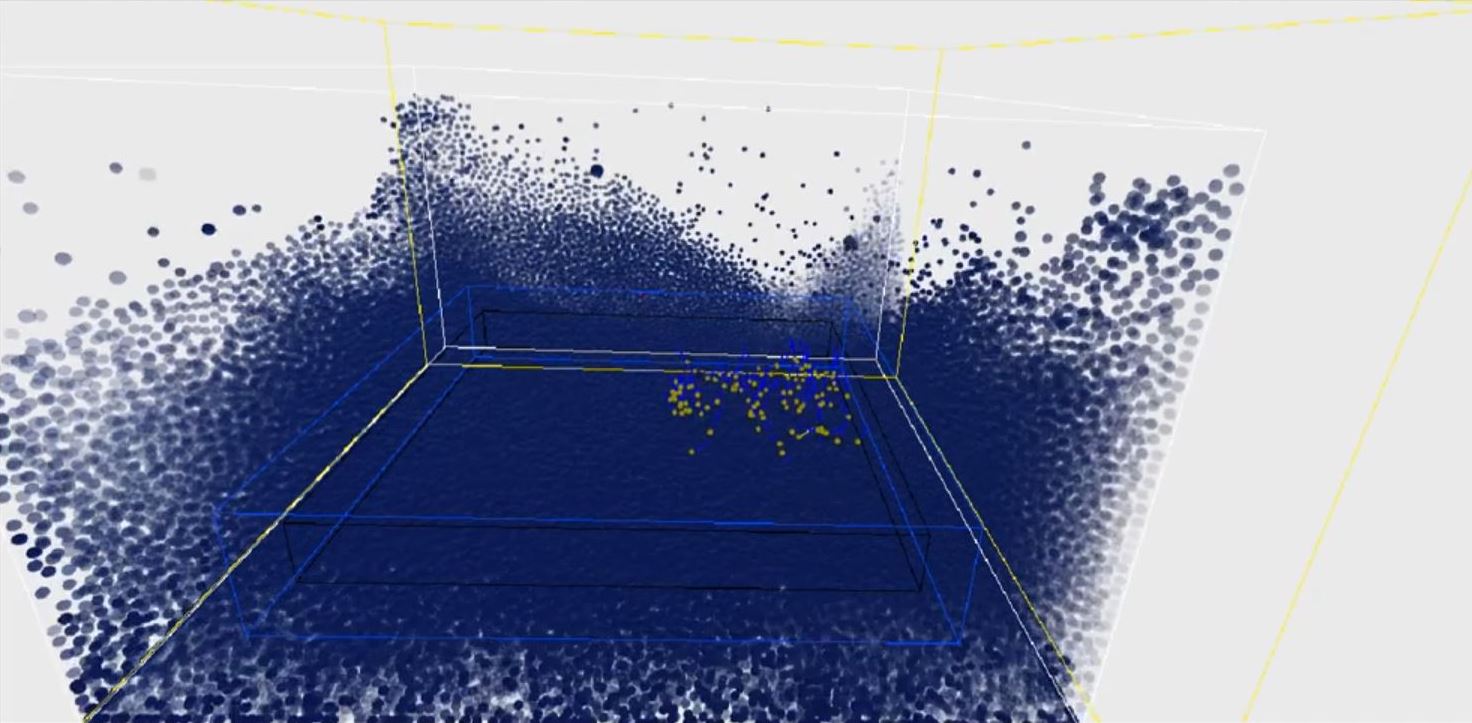}
\end{figure}
\begin{figure}[h!]
    \centering
    \includegraphics[width = 0.23\textwidth, height = 0.15\textwidth]{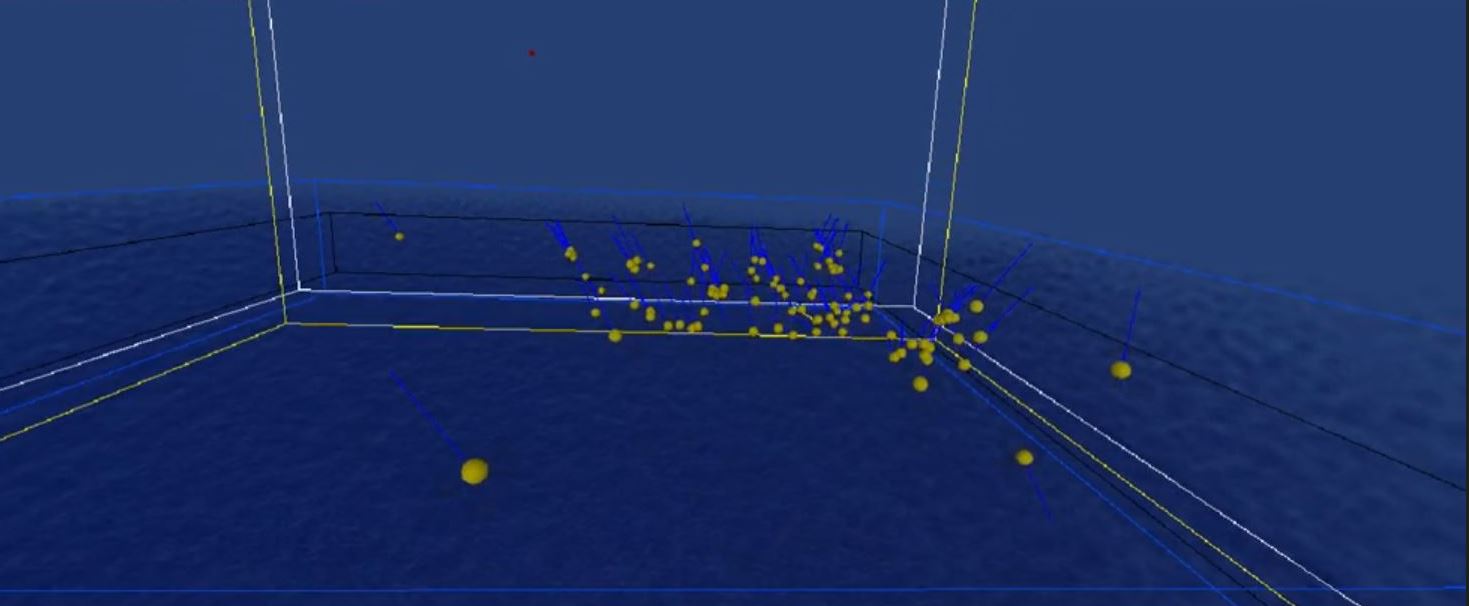}
    \includegraphics[width = 0.23\textwidth, height = 0.15\textwidth]{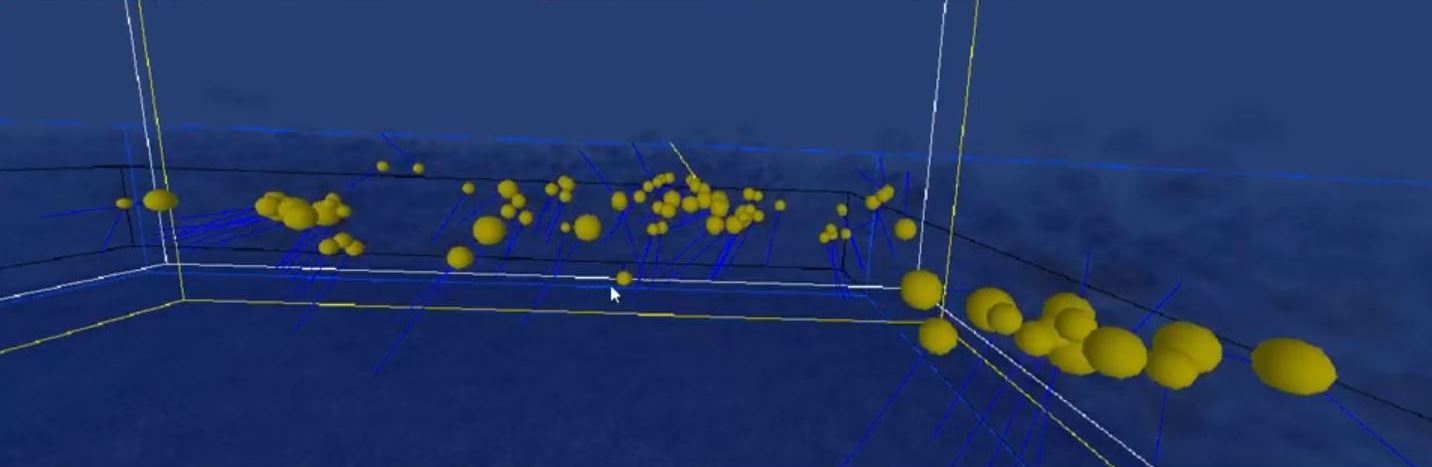}
\end{figure}
\begin{figure}[h!]
    \centering
    \includegraphics[width = 0.49\textwidth, height = 0.3\textwidth]{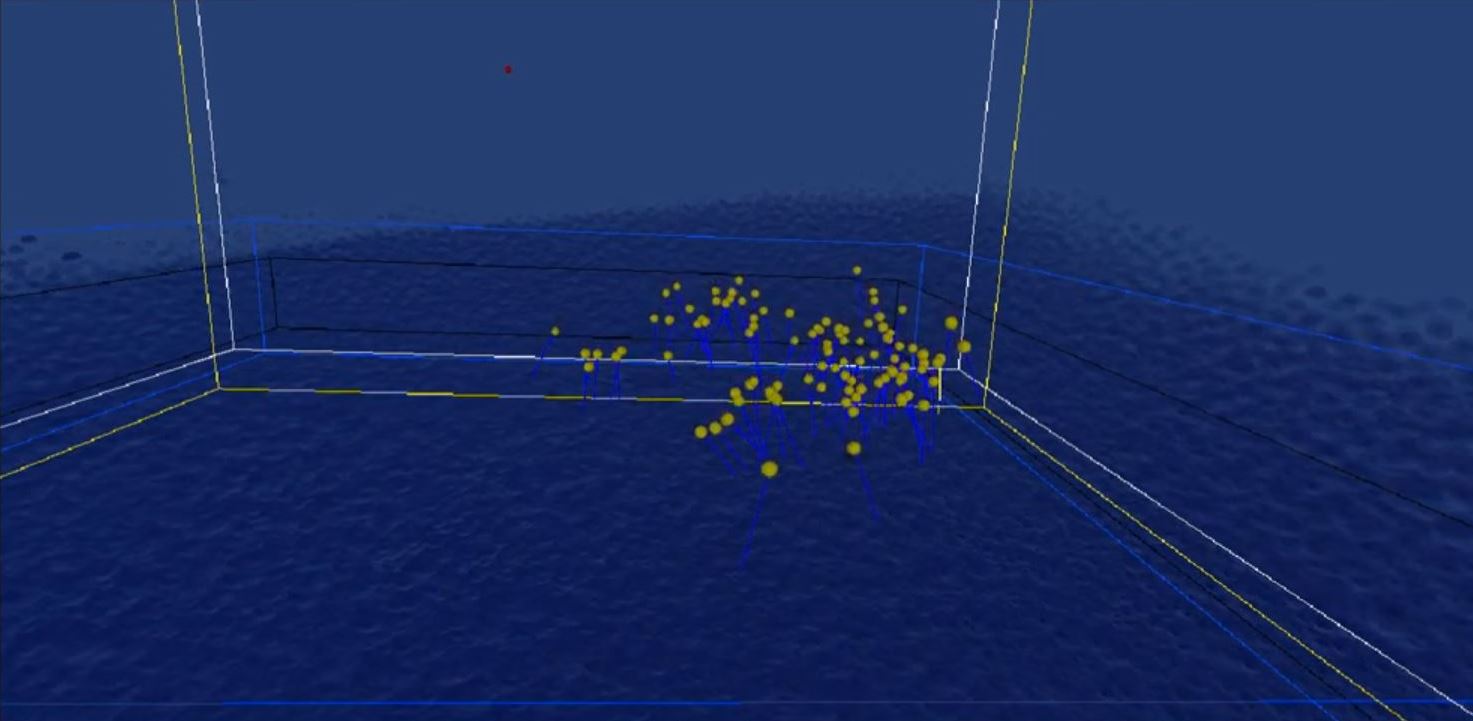}
    \caption{SPH boid interactions}
    \label{sph boid}
\end{figure}
\section{Conclusion}
We have developed a hydrodynamic simulation system using GPU compute shaders of DirectX that allows us to simulate virtual agent behaviors and navigation within a fluid environment based on smoothed particle hydrodynamics. The boid agents influence the motion of SPH fluid particles, while the forces from the SPH algorithm affect the movement of the boids. Our system performs real time water mesh surface reconstruction using a parallel marching cubes algorithm by calculating the voxel density values of each cube corner based on the signed distance field and then generating triangles that form the surface wrapping around the particles. Our system demonstrates its versatility by supporting reinforced robotic agents in addition to boid agents for underwater navigation and remote control engineering purposes. Our current research has provided valuable insights into the interactions between self-propelled agents, fluid dynamics, and the integration of smoothed particle hydrodynamics (SPH) with virtual boid agents. Looking ahead, there are several promising avenues for further exploration. As hardware technology advances, it would be beneficial to adapt our simulator to run on more powerful GPUs, such as the A100 or H100, to take advantage of their enhanced capabilities and potentially achieve even more complex and detailed simulations. We also see opportunities for applying our platform to research in microfluidics, nanofluidics, and the bioorganic realm, further extending its scope and relevance in these domains.

\section{Declaration of competing interest}
The authors declare that they have no known competing financial interests or personal relationships that could have appeared
to influence the work reported in this paper.





\bibliographystyle{elsarticle-num}
\bibliography{references}







\end{document}